\documentclass[11pt, oneside]{article}
\usepackage{jheppub}
\usepackage{amsmath}
\usepackage{amssymb}
\usepackage{amsfonts}
\usepackage{amsthm}
\usepackage{amsbsy}
\usepackage{array}
\usepackage{mathtools}
\usepackage[usenames,dvipsnames]{xcolor}
\usepackage{caption}
\usepackage{subcaption}
\usepackage{dsdshorthand}
\usepackage{graphicx}
\usepackage[vcentermath]{youngtab}
\usepackage{tikz}
\usepackage{booktabs,multirow}
\usepackage{listings}
\usepackage{xurl}
\usepackage{hyperref}

\definecolor{darkblue}{rgb}{0,0,0.5}
\definecolor{darkgreen}{rgb}{0,0.3,0}
\definecolor{darkpink}{rgb}{0.4,0,0.3}
\definecolor{graygreen}{rgb}{0.3,0.5,0.3}
\definecolor{grayblue}{rgb}{0.2,0.2,0.6}
\definecolor{grayred}{rgb}{0.5,0.2,0.2}

\newtheorem{hypothesis}{Hypothesis}

\makeatletter
\def\@fpheader{\ }
\makeatother

\title{Accurate bootstrap bounds from optimal interpolation}
\author{Cyuan-Han Chang$^{1,2}$, Vasiliy Dommes$^1$, Petr Kravchuk$^3$, David Poland$^4$, David Simmons-Duffin$^1$}
\affiliation{$^1$Walter Burke Institute for Theoretical Physics, Caltech, Pasadena, California 91125, USA}
\affiliation{$^2$Leinweber Institute for Theoretical Physics, University of Chicago, Chicago, Illinois 60637, USA}
\affiliation{$^3$Department of Mathematics, King's College London, Strand, London, WC2R 2LS, UK}
\affiliation{$^4$Department of Physics, Yale University, 217 Prospect St, New Haven, CT 06520, USA}
\emailAdd{cchang10@uchicago.edu}
\emailAdd{vdommes@caltech.edu}
\emailAdd{petr.kravchuk@kcl.ac.uk}
\emailAdd{david.poland@yale.edu}
\emailAdd{dsd@caltech.edu}

\date{}
\abstract{We develop new methods for approximating conformal blocks as positive functions times polynomials, with applications to the numerical bootstrap. We argue that to obtain accurate bootstrap bounds, conformal block approximations should minimize a certain error norm related to the asymptotics of dispersive functionals. This error norm can be made small using interpolation nodes with an appropriate optimal density. The optimal density turns out to satisfy a kind of force-balance equation for charges in 1-dimension, which can be solved using standard techniques from large-$N$ matrix models. We also describe how to use optimal density interpolation nodes to improve condition numbers inside the semidefinite program solver {\tt SDPB}. Altogether, our new approximation scheme and improvements to condition numbers lead to more accurate bootstrap bounds with fewer computational resources. They were crucial in the recent bootstrap study of stress tensors in the 3d Ising CFT.}

\preprint{CALT-TH 2025-031}

\begin{document}

\maketitle
\pagenumbering{roman}
\setcounter{page}{2}
\newpage
\pagenumbering{arabic}
\setcounter{page}{1}

\setlength{\parskip}{1pt}

\section{Introduction}
\label{sec:intro}

Bootstrap computations explore the intricate interplay between symmetries and positivity. In conformal field theory (CFT), symmetries give rise to conformal blocks --- special functions that encode the contribution of a conformal representation to a correlation function. In general, conformal blocks cannot be written in terms of elementary functions, so to work with them numerically we must approximate them in some way. In applications of semidefinite programming (SDP) to the conformal bootstrap \cite{Poland:2011ey,Kos:2013tga,Kos:2014bka,Simmons-Duffin:2015qma,Landry:2019qug}, one typically approximates conformal blocks in terms of polynomials of $\Delta$.

In our recent work on bootstrapping stress tensors in the 3d Ising CFT \cite{Chang:2024whx}, we ran up against limitations of previous methods for approximating conformal blocks. We found that, upon increasing the derivative order of our bootstrap calculations, we were forced to dramatically increase the degree of polynomials in our conformal block approximations. (Failure to do so led to nonsensical results like ostensibly ruling out the 3d Ising CFT.) We also had to increase precision in order to accommodate higher condition numbers inside the SDP solver {\tt SDPB} \cite{Simmons-Duffin:2015qma,Landry:2019qug}. Altogether, these increases in polynomial degree and precision introduced huge performance penalties. In short, our polynomial approximations were not efficient enough.

In this work, we introduce better methods for approximating conformal blocks in terms of polynomials, with applications to the numerical bootstrap. These methods have been crucial for obtaining high-derivative order bootstrap bounds on the 3d Ising CFT in \cite{Chang:2024whx,Chang:2025toappear}. Our first key observation, in section~\ref{sec:errornorm}, is that one shouldn't simply approximate the blocks themselves, but rather the action of the optimal functional $\alpha_\mathrm{opt}(F_{\De,j})$ on conformal blocks, since this is what determines the corresponding bootstrap bound. We make an educated guess that a typical optimal functional behaves like a ``dispersive" functional \cite{Mazac:2019shk,Penedones:2019tng,Caron-Huot:2020adz,Carmi:2020ekr} for the most important range of $\Delta$, and use this to define an error norm $\|\cdot\|_\mathrm{bootstrap}$ that quantifies ``goodness of approximation" for numerical bootstrap purposes.

The task of minimizing the error norm $\|\cdot\|_\mathrm{bootstrap}$ is a problem in optimal polynomial interpolation. In section~\ref{sec:alltheinterpolation}, we show that there exists an optimal density of interpolation nodes which ensures small errors with respect to $\|\cdot\|_\mathrm{bootstrap}$. This optimal density satisfies a kind of 1-dimensional charge-balance equation that appears in many physical contexts, including large-$N$ single matrix models. We compute it by adapting familiar techniques from the matrix model context.

In section~\ref{sec:conditionnumbers}, we describe how optimal density interpolation nodes can be used to make other choices for setting up Polynomial Matrix Programs (PMPs) that improve condition numbers of internal matrices in the SDP solver. Improved condition numbers mean that the solver can be run at lower precision, which helps performance. We also describe details of our implementation of these choices in {\tt SDPB}.

In section~\ref{sec:numericaltests}, we compare our new conformal block approximations and choices for setting up PMPs to previous methods. We find that one can obtain bootstrap bounds with significantly higher accuracy using smaller SDPs (as quantified by their ``primal dimension"). Furthermore, accuracy can be exponentially improved by simply increasing the number of interpolation nodes. We also demonstrate dramatic improvement in condition numbers inside {\tt SDPB}, and observe that condition numbers are essentially insensitive to increasing the number of interpolation nodes. We conclude with discussion and future directions in section~\ref{sec:future}.

\section{Quantifying approximation errors in the numerical bootstrap}
\label{sec:errornorm}

Numerical conformal bootstrap computations are optimization problems that involve the action of linear functionals on conformal blocks. As a simple example, consider bootstrapping a four-point function of identical scalar operators $\<\f(x_1)\f(x_2)\f(x_3)\f(x_4)\>$ \cite{Rattazzi:2008pe}. A commonly used type of functional consists of linear combinations of derivatives around the point $(z,\bar z)=(\frac 1 2,\frac 1 2)$ in cross ratio space:
\be
\label{eq:derivativebasis}
\a(F) &\equiv \sum_{\substack{m+n \textrm{ odd} \\ m+n\leq \Lambda}}\left.a_{mn}\ptl_z^m \ptl_{\bar z}^n F(z,\bar z)\right|_{(z,\bar z) = (\frac 1 2,\frac 1 2)}.
\ee
Here, $a_{mn}$ are decision variables that we must determine using a convex optimization solver. The functional $\a$ is typically constrained by positivity conditions like
\be
\label{eq:positivityconditions}
\a(F_{\De,j}) \geq 0 \qquad \textrm{for }j=0,2,\dots,j_\mathrm{max}\textrm{ and }\De\in[\De_\mathrm{min}(j),\oo),
\ee
where
\be
F_{\De,j}(z,\bar z) &= (z\bar z)^{-\De_\f} g_{\De,j}(z,\bar z) - \Big((z,\bar z) \leftrightarrow (1-z,1-\bar z)\Big),
\ee
and $g_{\De,j}(z,\bar z)$ are conformal blocks. We will choose conventions where conformal blocks have the small $z,\bar z$ behavior
\be
\label{eq:ourblockconvention}
g_{\De,j}(z,\bar z) &\sim (z \bar z)^{\frac{\De}{2}} \frac{j!}{(-2)^j(\frac{d-2}{2})_j} C^{\frac{d-2}{2}}_j\p{\frac{z+\bar z}{2\sqrt{z \bar z}}}\qquad (z,\bar z\to 0),
\ee
where $C^{\frac{d-2}{2}}_j(\cos\th)$ is a Gegenbauer polynomial.\footnote{\label{foot:structures}These are the same conventions used in~\cite{Poland:2018epd}. Blocks with this normalization arise from fusing two standard conformal three-point structures of the form
\be
\frac{Z_3^{\mu_1}\cdots Z_3^{\mu_j}-\textrm{traces} }{|x_{12}|^{\De_1+\De_2-\De_3}|x_{23}|^{\De_2+\De_3-\De_1}|x_{31}|^{\De_3+\De_1-\De_2}},\qquad Z_3^\mu = \frac{|x_{23}||x_{31}|}{|x_{12}|} \p{\frac{x_{13}^\mu}{x_{13}^2} - \frac{x_{23}^\mu}{x_{23}^2}},
\ee
using a standard conformal two-point structure, by summing over descendants. In general, for spinning blocks we adopt conventions where the blocks are obtained by fusing three-point structures of the form $V^{abc}(x_1,x_2,x_3)/\p{|x_{12}|^{\De_1+\De_2-\De_3}|x_{23}|^{\De_2+\De_3-\De_1}|x_{31}|^{\De_3+\De_1-\De_2}}$, where $V^{abc}(x_1,x_2,x_3)$ is a $\De$-independent tensor carrying indices for the $\SO(d)$ representations of the three operators. These are the conventions used by, e.g.\ {\tt blocks\_3d} \cite{Erramilli:2020rlr}.} We typically seek to maximize some $\a$-dependent objective over all $\a$ satisfying (\ref{eq:positivityconditions}).

In order to encode the positivity conditions (\ref{eq:positivityconditions}) on a computer, we must approximate the conformal blocks and their derivatives $\ptl_z^m \ptl_{\bar z}^n g_{\De,j}(\frac 1 2,\frac 1 2)$ in some way as a function of $\Delta$ (for each $j$). How can we determine how ``good" our approximation is?

One issue is that the normalization of conformal blocks is ambiguous. Someone else might choose to normalize their conformal blocks differently from ours: $g^{\textrm{theirs}}_{\De,j} = f_j(\Delta) g^{\textrm{ours}}_{\Delta,j}$, where the $f_j(\Delta)$ are positive functions of $\Delta$.\footnote{See Table I of~\cite{Poland:2018epd} for a variety of choices that have been used in the literature.} Such a change of normalization can be reabsorbed into a redefinition of OPE coefficients, so there is no obvious physical reason to prefer one normalization over another. Furthermore, the positivity conditions $\a(F_{\Delta,j})\geq 0$ are blind to the choice of normalization. However, the choice of normalization matters when we consider approximations to the blocks. For example, we might measure the distance between a function $g(\De)$ and its approximation $g_\mathrm{approx}(\De)$ using the uniform norm $\|g-g_\mathrm{approx}\|=\sup_{\Delta} |g(\Delta)-g_\mathrm{approx}(\Delta)|$. If we re-weight both functions by some $f(\Delta)$, the distance can potentially change by a large amount. Is there a ``good" choice of normalization that leads to the best approximations for numerical bootstrap purposes?

Another issue is: which derivatives of conformal blocks should we be most careful to approximate well? Should we focus on zeroth derivatives or 20th derivatives?

To resolve these questions, let us take a step back and ask what we {\it really\/} want to approximate. Ultimately, we care about the accuracy of the resulting bootstrap bound. We want our bound to be as close as possible to what it would have been if we used exact expressions for conformal blocks. The value of a bootstrap bound is determined by the corresponding optimal/extremal functional $\a_\mathrm{opt}$ (computed by the convex optimization solver). This suggests the following hypothesis:
\begin{hypothesis}
\label{hyp:chicken}
To compute a bootstrap bound as accurately as possible, we should approximate the action of the optimal functional $\a_\mathrm{opt}(F_{\De,j})$ as accurately as possible as a function of $\Delta$ (for each $j$).
\end{hypothesis}
If we have a good approximation for $\a_\mathrm{opt}(F_{\De,j})$, then we will also have good approximations for the locations of its zeros --- i.e.\ the positivity conditions that are saturated at the optimum, as well as for the value of the objective. Hypothesis~\ref{hyp:chicken} resolves the ambiguity of whether we should try to approximate the block or its derivatives: we should try to approximate the particular linear combination of derivatives given by $\a_\mathrm{opt}(F_{\De,j})$.\footnote{Of course, it is also important that our approximation for the action of other (non-optimal) functionals be reasonably accurate as well, so that we don't create spurious solutions to the convex optimization problem far away from the correct optimum.}

Unfortunately, hypothesis~\ref{hyp:chicken} leads to a chicken-and-egg problem: how can we approximate $\a_\mathrm{opt}(F_{\De,j})$ before we know what $\alpha_\mathrm{opt}$ is? In order to know $\a_\mathrm{opt}$, we must run the solver, but in order to set up the solver we need to know $\a_\mathrm{opt}$. 

Fortunately, there is at least one bootstrap bound in $d>1$ where the optimum is known analytically, and we can use this example to build some intuition. This is the ``spin-2 gap" problem: maximize the dimension $\De_2$ of the lowest-dimension spin-2 operator in the $\f\x\f$ OPE, as a function of $\De_\f$. This bound is saturated by the GFF value $\De_2^\mathrm{GFF}=2\De_\f + 2$, and a corresponding optimal functional is known analytically \cite{Caron-Huot:2020adz}. The optimal functional is a so-called ``dispersive" functional with Regge spin $k=2$. Dispersive functionals are certain integrals of conformal blocks over two copies of the cut plane $\C-((-\oo,0]\cup [1,\oo))$ (one copy for each cross ratio $z,\bar z$). The Regge spin $k=2$ corresponds to the most general kind of dispersive functional that is ``swappable" \cite{Qiao:2017lkv} when acting on the crossing equation. A dispersive functional $\Phi_k$ with Regge spin $k$ acts in the following way on blocks at large $\Delta$ \cite{Caron-Huot:2021enk,Chang:2023szz}:
\be
\Phi_k(F_{\De,j}) &\sim 4^{\De} \De^{\frac d 2 - 2(k-1) - 4 \De_\f} \sin\p{\pi \tfrac{\tau-2\De_\f}{2}}^2 \qquad (\textrm{large $\De$},\  \textrm{fixed $j$}),
\label{eq:dispersivefunctionalasymptotics}
\ee
where the conformal blocks are normalized according to (\ref{eq:ourblockconvention}), and $\tau=\De-j$ is the twist. Thus, dispersive functionals oscillate with $\De$ and grow at a particular rate dictated by (\ref{eq:dispersivefunctionalasymptotics}).

\begin{figure}
\centering
\begin{subfigure}[b]{0.49\textwidth}
    \includegraphics[width=\textwidth]{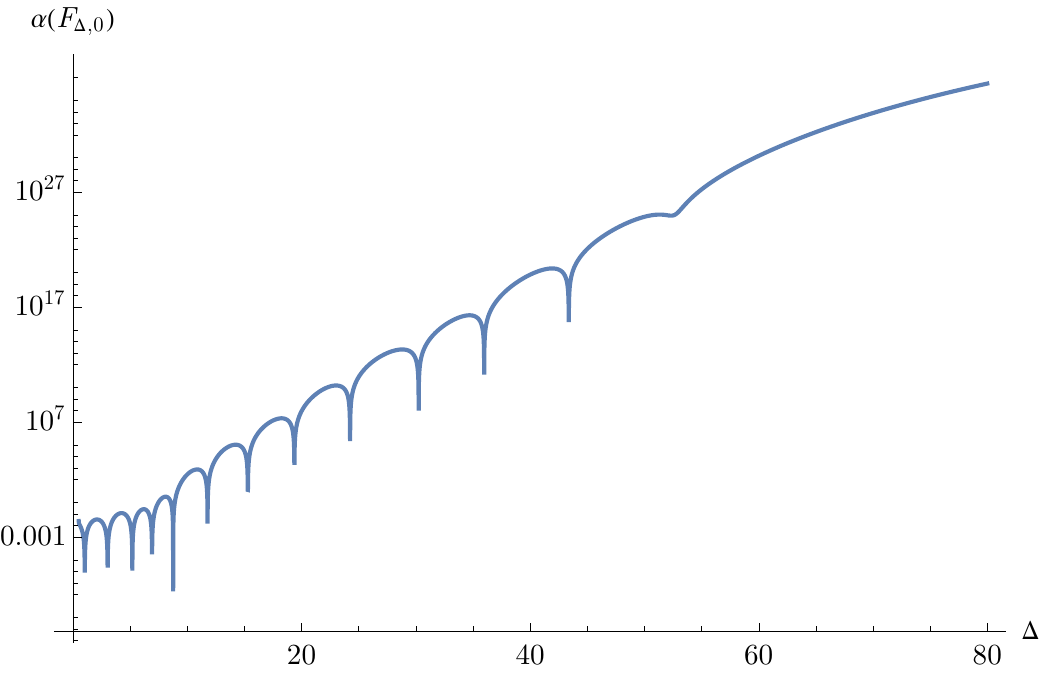}
    \caption{unnormalized $\a(F_{\De,0})$}
    \label{fig:file1}
  \end{subfigure}
  \hfill
  \begin{subfigure}[b]{0.49\textwidth}
    \includegraphics[width=\textwidth]{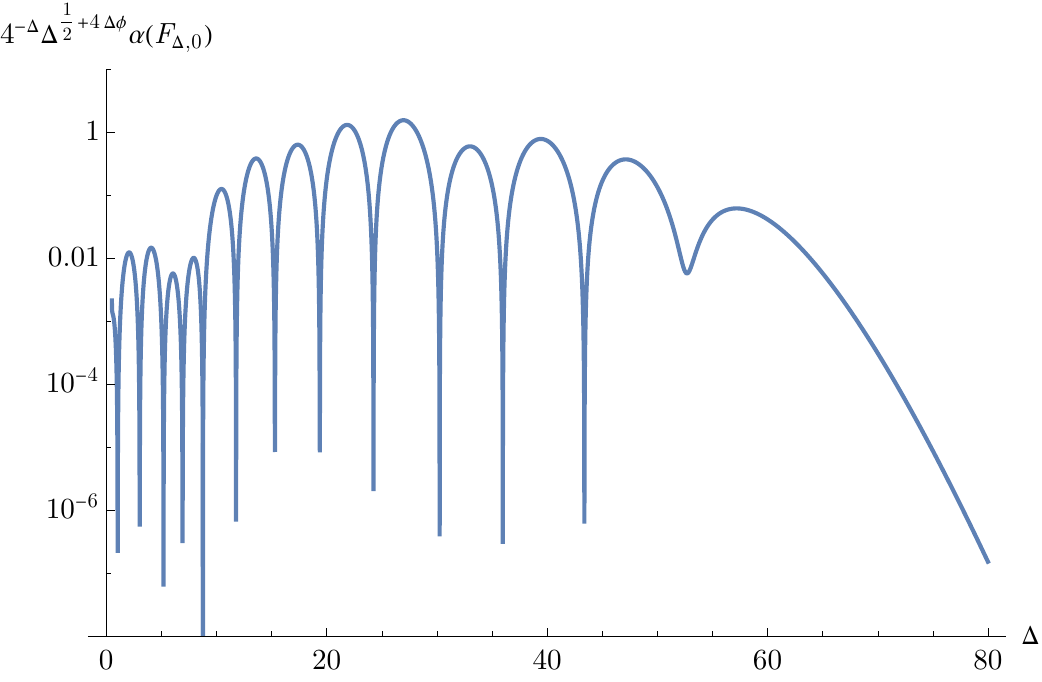}
    \caption{normalized $4^{-\De}\De^{\frac 1 2 + 4\De_\f} \a(F_{\De,0})$}
    \label{fig:file2}
  \end{subfigure}
\caption{\label{fig:spin2gap}The optimal functional for the ``spin-2 gap" problem in 3d with $\De_\f=0.5181489$, acting on conformal blocks with $j=0$. In figure~\ref{fig:file1}, we show the action of the functional $\a(F_{\De,0})$ on blocks with the normalization convention (\ref{eq:ourblockconvention}). In figure~\ref{fig:file2}, we show the ``normalized" action given by dividing by the asymptotics of a spin-2 dispersive functional $4^{-\De}\De^{\frac 1 2 + 4\De_\f} \a(F_{\De,0})$. The normalized action remains $O(1)$ in the region $\De\lesssim 60$ where the functional exhibits interesting structure, before exponentially decaying at large $\De$.}
\end{figure}

For numerical applications, we are interested in the behavior of functionals with a finite number of derivatives $\Lambda$. So let us study the optimal functional for the spin-2 gap problem with large but finite $\Lambda$. In figure~\ref{fig:spin2gap}, we show plots of the optimal functional for the spin-2 gap problem computed at derivative order $\Lambda=59$ for $\De_\f=0.5181489$, acting on a conformal block with $j=0$. In figure~\ref{fig:file1} we display the ``unnormalized" functional $\a(F_{\De,0})$, which varies by over 40 orders of magnitude for the displayed range of $\De$. After exhibiting some nontrivial structure for $\De\lesssim 60$, it continues to grow before reaching a maximum of about $10^{40}$ near $\De=150$, and eventually decaying exponentially. This is clearly not a good normalization for computing approximations --- the functional is exponentially large in the regime of $\De$ where its behavior is least interesting.

By contrast, in figure~\ref{fig:file2}, we display the functional normalized by dividing by the large-$\De$ asymptotics of a spin-2 dispersive functional (omitting the $\sin^2(\pi \frac{\tau-2\De_\f}{2})$ factor, since it would create singularities):
\be
\label{eq:normalizedfunctional}
4^{-\De}\De^{\frac 1 2 + 4\De_\f} \a(F_{\De,0}).
\ee
The normalized functional oscillates for a long time as a function of $\Delta$, with its amplitude remaining remarkably stable, before exponentially decaying for $\De\gtrsim 60$. We see that (\ref{eq:normalizedfunctional}) is an excellent normalization for the optimal functional: it remains $O(1)$ for precisely the range of $\Delta$ where it exhibits interesting structure (such as the double zeros characteristic of extremal functionals). The eventual exponential decay in figure~\ref{fig:file2} is a consequence of our choice of the derivative basis (\ref{eq:derivativebasis}), together with the large-$\De$ behavior of conformal blocks \cite{Pappadopulo:2012jk,Hogervorst:2013sma}:
\be
g_{\De,j}(z,\bar z) &\overset{\De\to\oo}{\sim} (4r)^{\De} \x \textrm{finite},\qquad r=\sqrt{\rho\bar \rho},
\ee
where $\rho=z/(1+\sqrt{1-z})$ and $\bar \rho=\bar z/(1+\sqrt{1-\bar z})$ are the radial coordinates of \cite{Hogervorst:2013sma}. At the crossing-symmetric point $z=\bar z = \frac 1 2$, we have $r=r_*\equiv 3-2\sqrt 2 \approx 0.17$. Consequently, any finite linear combination of conformal blocks and their derivatives around this point will behave like $(4r_*)^\De$ at large $\De$. In particular, the combination $4^{-\De}\De^{\frac 1 2 + 4\De_\f} \a(F_{\De,0})$ decays exponentially like $r_*^\De$ for sufficiently large $\De$. The $O(1)$ oscillations before the exponential decay are a consequence of a delicate balance between polynomial-like contributions from taking derivatives of the block and the exponential factor $r_*^\Delta$.

The ``spin-2 gap" example leads to our second hypothesis:
\begin{hypothesis}
\label{hyp:crazy}
For a generic numerical bootstrap bound, a good normalization for the functional is obtained by dividing by the asymptotics of a spin-2 dispersive functional.
\end{hypothesis}
Here, ``good" means that it remains $O(1)$ in the range of $\De$ where the functional has interesting structure, before exponentially decaying.
Hypothesis~\ref{hyp:crazy} is reasonable, but speculative given our limited knowledge of optimal bootstrap bounds. However, we can test it numerically in other cases.

As another example, let us consider an optimal functional for an OPE coefficient bound in the mixed correlator system of $T,\sigma,\epsilon$ in the 3d Ising CFT \cite{Chang:2024whx}. In a mixed-correlator system, $\a(F_{\De,j})$ will be a matrix whose elements $\a(F_{\De,j})_{ij,kl}$ are paired with products of OPE coefficients $\l_{\cO_i\cO_j\cO_{\De,j}}\l_{\cO_k\cO_l\cO_{\De,j}}$. For a spin-2 dispersive functional in $d=3$, the diagonal entries of this matrix will behave like
\be
\Phi_2(F_{\De,j})_{ij,ij} &\sim f(\De_i,\De_j,\De) \sin\p{\pi \tfrac{\tau-\De_i-\De_j}{2}}^2,\quad \textrm{where}\nn\\
f(\De_i,\De_j,\De) &= 4^{\De+\De_i+\De_j}\De^{-\frac 1 2 - 2\De_i-2\De_j} \G(\De_i)\G(\De_i-\tfrac 1 2) \G(\De_j)\G(\De_j-\tfrac 1 2).
\ee
This is a more refined version of (\ref{eq:dispersivefunctionalasymptotics}) for mixed blocks, where we have kept track of $\De_i$ and $\De_j$ dependence. (This formula holds for blocks with spinning external operators as well, provided we use the conventions described in footnote~\ref{foot:structures}.) Thus, hypothesis~\ref{hyp:crazy} suggests that we should normalize the functional by dividing each matrix element $\a(F_{\De,j})_{ij,kl}$ by $\sqrt{f(\De_i,\De_j,\De)f(\De_k,\De_l,\De)}$ (note that this transformation preserves positive-definiteness, since it is equivalent to a rescaling of OPE coefficients).

\begin{figure}
\centering
\begin{subfigure}[b]{0.49\textwidth}
    \includegraphics[width=\textwidth]{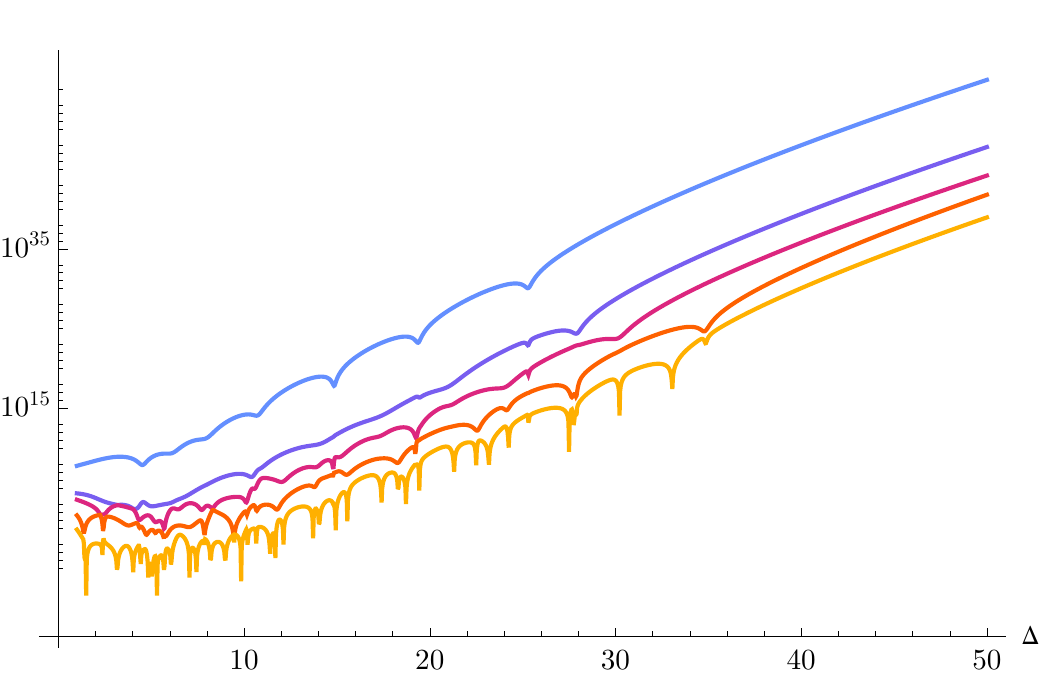}
    \caption{unnormalized}
    \label{fig:file1tse}
  \end{subfigure}
  \hfill
  \begin{subfigure}[b]{0.49\textwidth}
    \includegraphics[width=\textwidth]{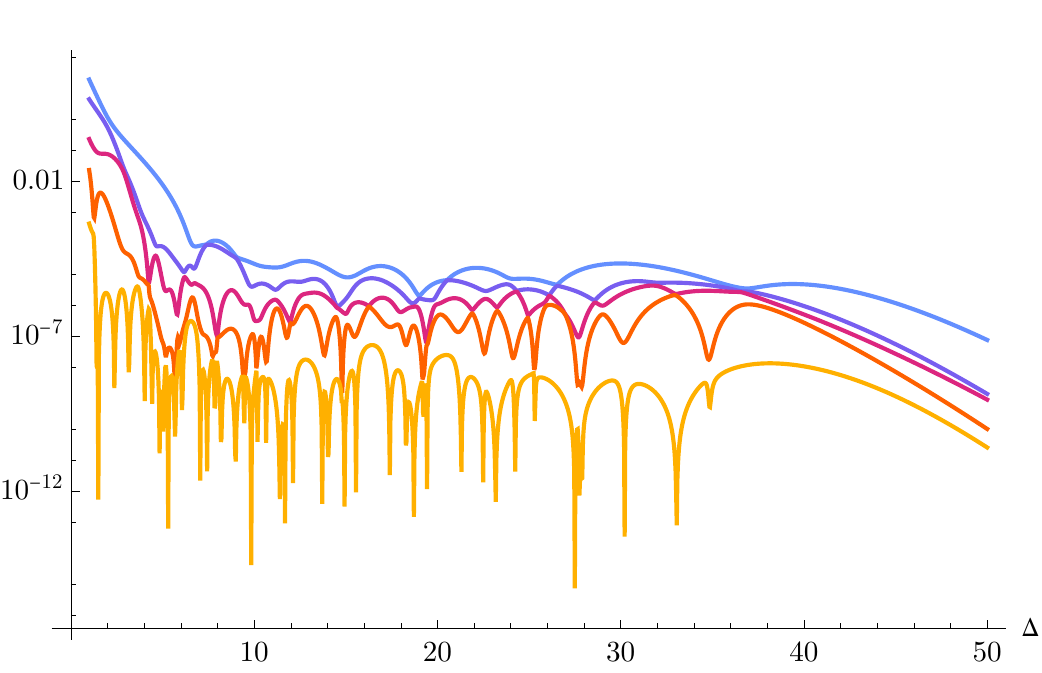}
    \caption{normalized}
    \label{fig:file2tse}
  \end{subfigure}
\caption{\label{fig:tse}The optimal functional for an OPE coefficient bound in the $T,\sigma,\epsilon$ system in the 3d Ising CFT, computed with $\Lambda=43$. We choose $\De_\s,\De_\e$ inside the allowed island, and choose the OPE coefficient vector $\vec \l = (\l_B,\l_F,\l_{\s\s\e},\l_{\e\e\e},\l_{TT\e})$ to point in an allowed direction, and maximize its magnitude $|\vec \l|$. We plot the eigenvalues of the resulting optimal functional, acting on the contribution of parity-even $\Z_2$-even operators with spin $j=4$. The eigenvalues are positive by construction. In figure~\ref{fig:file1tse}, we show the eigenvalues of the unnormalized functional action. In figure~\ref{fig:file2tse} we show the eigenvalues of the normalized matrix obtained by dividing each matrix element by factors $\sqrt{f(\De_i,\De_j,\De)f(\De_k,\De_l,\De)}$ coming from the large-$\De$ asymptotics of Regge spin-2 dispersive functionals.}
\end{figure}

In figure~\ref{fig:tse}, we plot the optimal functional for the $T,\sigma,\epsilon$ OPE bound computed with $\Lambda=43$, acting on parity-even $\Z_2$-even operators with spin 4. There are 5 possible three-point structures between such operators and pairs of $T,\sigma,$ or $\e$, so $\a(F_{\De,4})$ is a $5\x5$ matrix.  Figure~\ref{fig:file1tse} shows eigenvalues of the unnormalized matrix $\a(F_{\De,4})$. The unnormalized matrix has eigenvalues that differ by 20 orders of magnitude and grow exponentially into the ``uninteresting" region of large $\De$. By contrast, in figure~\ref{fig:file2tse}, we show the eigenvalues of the normalized matrix obtained by dividing each element by $\sqrt{f(\De_i,\De_j,\De)f(\De_k,\De_l,\De)}$. The normalized matrix has eigenvalues that differ by only 5 orders of magnitude and remain stable in the ``interesting" region $\De\lesssim 40$ before exponentially decaying. 

Before proceeding, let us describe an alternative motivation for how to normalize functionals that avoids the speculative connection to dispersive functionals, but leads to similar conclusions. One natural way to normalize conformal blocks is to multiply them by OPE coefficients of a standard solution to crossing symmetry. Mean Field Theory (MFT) gives such a solution: the MFT four-point function is a sum of $p^\mathrm{MFT}_{\De,j} g_{\De,j}$, where the squared OPE coefficients $p^\mathrm{MFT}_{\De,j}$ are known analytically \cite{Fitzpatrick:2011dm}. (They are defined for $\De=2\De_\f+2n+j$ for $n\in \Z_{\geq 0}$, and we choose an analytic continuation in $\De$ that is well-behaved in the right-half $\De$ plane.) The combination $p^\mathrm{MFT}_{\De,j} g_{\De,j}$ is then convention-independent, and measures essentially how much a given conformal block contributes to the four-point function of MFT\@. The lightcone bootstrap suggests that every four-point function behaves like MFT in a certain kinematic limit \cite{Komargodski:2012ek,Fitzpatrick:2012yx}, so we might guess that $p^\mathrm{MFT}_{\De,j} g_{\De,j}$ is a reasonable guess for the rough contribution of a block to a four-point function. Similarly, \cite{Mukhametzhanov:2018zja} argues that large-$\De$ asymptotics of OPE coefficients at fixed $j$ in general CFTs agree with those of MFT\@. So perhaps we should multiply $\alpha(F_{\De,j})$ by $p_{\De,j}^\mathrm{MFT}$?

The asymptotics of scalar MFT OPE coefficients (at fixed $j$) are
\be
\label{eq:mftasympt}
p_{\De,j}^\mathrm{MFT} &\sim 4^{-\De}\De^{4\De_\f-\frac{3d}{2}}, \qquad  (\textrm{large $\De$},\  \textrm{fixed $j$}).
\ee
We see that (\ref{eq:mftasympt}) differs from the reciprocal of the $k=2$ dispersive functional asymptotics (\ref{eq:dispersivefunctionalasymptotics}) by a factor of $\De^{-2-d}$. Thus, a plot of $p_{\De,j}^\mathrm{MFT} \alpha(F_{\De,j})$ would look similar to figures~\ref{fig:file2} and~\ref{fig:file2tse}, but with additional decay like $\De^{-5}$. This would produce a less ``flat" plot, but the resulting normalization would still work reasonably well with the interpolation method we discuss below. Importantly, this line of argument produces the same $4^{-\De}$ factor that we got from considering dispersive functionals.

Let us now discuss how to practically build approximations to conformal blocks, given our observations above. As is common in numerical bootstrap computations, we will approximate conformal blocks (and their derivatives) in terms of ``damped rationals," given by a product of an exponential term $(4r_*)^\De$ and a rational function:
\be
\left.\ptl_z^m\ptl_{\bar z}^n g_{\De,j}(z,\bar z)\right|_{z=\bar z = \frac 1 2} &\sim (4r_*)^\De \frac{P^{m,n}_j(\De)}{\prod_{i=1}^{n_\mathrm{poles}} (\De-\De_i)}.
\ee
Here, $\De_i$ runs over a subset of size $n_\mathrm{poles}$ of the full set of poles of a conformal block \cite{Kos:2013tga,Kos:2014bka,Penedones:2015aga,Erramilli:2019njx}. The subset we choose will affect the accuracy of approximation: by using larger $n_\mathrm{poles}$ in the denominator and higher degree polynomials in the numerator, we will be able to make more accurate approximations.

According to our hypotheses, we should find an accurate approximation for the  block divided by the asymptotics of a spin-2 dispersive functional $4^\De\De^{\frac d 2 - 2 - 4\De_\f}$ (or perhaps multiplied by the asymptotics of MFT OPE coefficients $4^{-\De} \De^{4\De_\f-\frac{3d}{2}}$). Actually, for simplicity and convenience, we will just divide by the exponential factor $4^\De$, and not the power-law factors $\De^{\#}$. For moderate values of $\De_\f$, the exponential factor is much more important for obtaining good approximation accuracy.\footnote{Some numerical bootstrap references \cite{Kos:2013tga,Kos:2014bka,Kos:2015mba,Kos:2016ysd,Li:2017ddj,Behan:2016dtz} have used alternative normalizations for conformal blocks that include a $4^{-\De}$ factor: $g^\mathrm{alternative}_{\De,j}\sim 4^{-\De} g^\mathrm{ours}_{\De,j}$. The present discussion can be interpreted as saying that such conventions are actually much better than (\ref{eq:ourblockconvention}) for building approximations to the blocks via interpolation.} Another simplification is that dividing only by $4^\De$ treats different blocks in mixed-correlator systems uniformly, and so is easier to implement. We discuss some of these issues in section~\ref{sec:future}. Future studies (particularly with large external dimensions) may wish to include the power-law factors $\De^{\frac 1 2 + 4\De_\f}$ as well when determining error norms for approximation. The formalism we develop in section~\ref{sec:optimalnodes} is capable of incorporating these factors.

To summarize, we seek an approximation of the form
\be
4^{-\De} \left.\ptl_z^m\ptl_{\bar z}^n g_{\De,j}(z,\bar z)\right|_{z=\bar z = \frac 1 2} &\sim r_*^\De \frac{P^{m,n}_j(\De)}{\prod_{i=1}^{n_\mathrm{poles}} (\De-\De_i)}.
\label{eq:thingwewanttodo}
\ee
We must try to find polynomials $P^{m,n}_j(\De)$ such that the two sides of (\ref{eq:thingwewanttodo}) differ as little as possible in the sup norm.
We can rephrase this requirement slightly as follows. Let us define the positive function
\be
\label{eq:muforconformalblocks}
\mu^{(j)}(\De) &\equiv \frac{r_*^\De }{\prod_{i=1}^{n_\mathrm{poles}} (\De-\De_i)}.
\ee
We want to find polynomials $P^{m,n}_j(\De)$ that minimize
\be
\|P_j^{m,n}-\tl G_j^{m,n}\|_{\mathrm{bootstrap},j},
\ee
where the error norm $\|\cdot\|_{\mathrm{bootstrap},j}$ is defined by
\be
\label{eq:bootstrapnorm}
\|f\|_{\mathrm{bootstrap},j} &\equiv \sup_{\De\geq \De_\mathrm{min}(j)} \mu^{(j)}(\De) |f(\De)|,
\ee
and
\be\label{eq:interpolationtarget}
\tl G_j^{m,n}(\De) &= \frac{4^{-\De} \left.\ptl_z^m\ptl_{\bar z}^n g_{\De,j}(z,\bar z)\right|_{z=\bar z = \frac 1 2}}{\mu^{(j)}(\De)}.
\ee

In \eqref{eq:muforconformalblocks}, we choose to keep the $n_\mathrm{poles}$ rightmost poles of a conformal block in $\mu^{(j)}(x)$. The rationale for doing so is that it extends the analyticity domain of the functions~\eqref{eq:interpolationtarget} that we are interpolating away from the positive real axis. While we have not attempted to study this question rigorously, it is a well-known fact in polynomial interpolation that the size of analyticity domain determines the exponential rate of convergence~\cite{Walsh}. For example, replacing the rightmost poles of a block with a different set of poles in $\mu^{(j)}(x)$ would shrink the analyticity domain of \eqref{eq:interpolationtarget}, and we have tested that the interpolation indeed gets worse if we use different poles.

Our strategy will be to build $P^{m,n}_j$ as an interpolating polynomial for $\tl G_j^{m,n}$ through a set of sample points that depend on $\mu^{(j)}(\De)$. Using interpolation is important from the point of view of hypothesis 1: we wish to approximate well the linear combination $\a_\text{opt}(F_{\De,j})$ rather than the individual conformal block derivatives. Since interpolation is a linear procedure, we can apply it to individual derivatives without knowing the coefficients $a_{mn}$ in $\a_\text{opt}$.\footnote{We also tried using the Remez algorithm to find the best polynomial approximations to individual conformal block derivatives (which is a non-linear procedure), but found that this was prohibitively slow and we did not pursue it further.} Intuitively, from figures~\ref{fig:file2} and \ref{fig:file2tse}, we might guess that the optimal sample points for interpolation should cover the range of $\De$ where the functional oscillates, but they can be sparse in the regime of larger $\De$ (where the normalized functional decays exponentially). In the next section, we will develop tools to produce a natural set of sample points associated with the error norm (\ref{eq:bootstrapnorm}). They will indeed be consistent with this guess.

\section{Interpolation}
\label{sec:alltheinterpolation}

\subsection{Optimal polynomial interpolation}
\label{sec:interpolation}

Polynomial interpolation of degree $n$ on a subset $U\subseteq \R$ can be viewed as a linear operator mapping the continuous functions on $U$ to their interpolating polynomials,~i.e.
\be
	I_n:C(U)\to P_n,
\ee
where $C(U)$ is the space of continuous functions on $U$ and $P_n$ is the space of all real polynomials of degree $n$. The interpolation operator $I_n$ is fully defined by the set $\{x_i\}_{i=0}^n$ of $n+1$ interpolation nodes $x_i$. For any $f\in C(U)$ we have
\be
	(I_nf)(x)=\sum_{k=0}^n f(x_k)\prod_{i\neq k} \frac{x-x_i}{x_k-x_i}.
\ee

As discussed in the previous subsection, our goal is to minimize the interpolation error $\|I_nf - f\|$ on $U=[0,+\oo)$, where $\|\.\|=\|\.\|_\text{bootstrap}$ is the norm on $C(U)$ given by~\eqref{eq:bootstrapnorm} with $x=\De-\De_\mathrm{min}(j)$. In this subsection, we discuss in general how the choice of the nodes $x_i$ affects interpolation errors and explain what we mean by an ``optimal'' set of interpolation nodes. In the following subsections, we will determine the optimal choice of $x_i$ for the norm $\|\.\|_\text{bootstrap}$ on $U=[0,+\oo)$.

We mostly follow the survey~\cite{BrutmanLebesgue}, generalizing as appropriate for our situation. It is helpful to first discuss the simplest interpolation problem, where $U=[-1,1]$ and the error is measured by the uniform norm, $\|f\|=\sup_{x\in [-1,1]}|f(x)|$. Since this norm treats all points in $[-1,1]$ uniformly, an uninitiated reader might guess that the optimal choice of interpolation nodes is given by uniformly spaced $x_i$,
\be\label{eq:uniform}
	x_i = \frac{2i}{n}-1.
\ee
It turns out that such uniform interpolation nodes lead to a very poor interpolation operator $I_n$. A much better choice are the Chebyshev nodes
\be\label{eq:chebyshev}
	x_i = \cos\p{\frac{2i+1}{2(n+1)}\pi}.
\ee

\begin{figure}[t]
	\centering
	\includegraphics[width=0.48\textwidth]{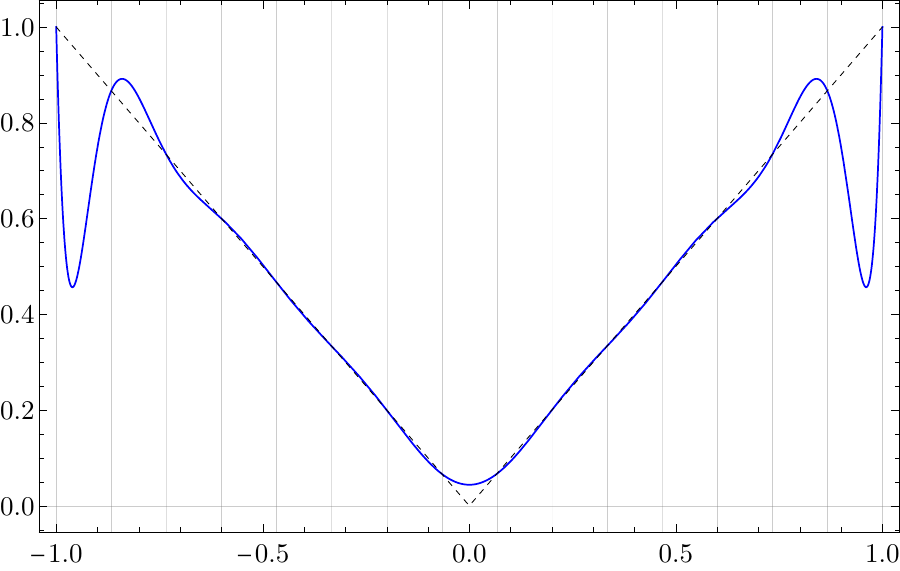}
~
	\includegraphics[width=0.48\textwidth]{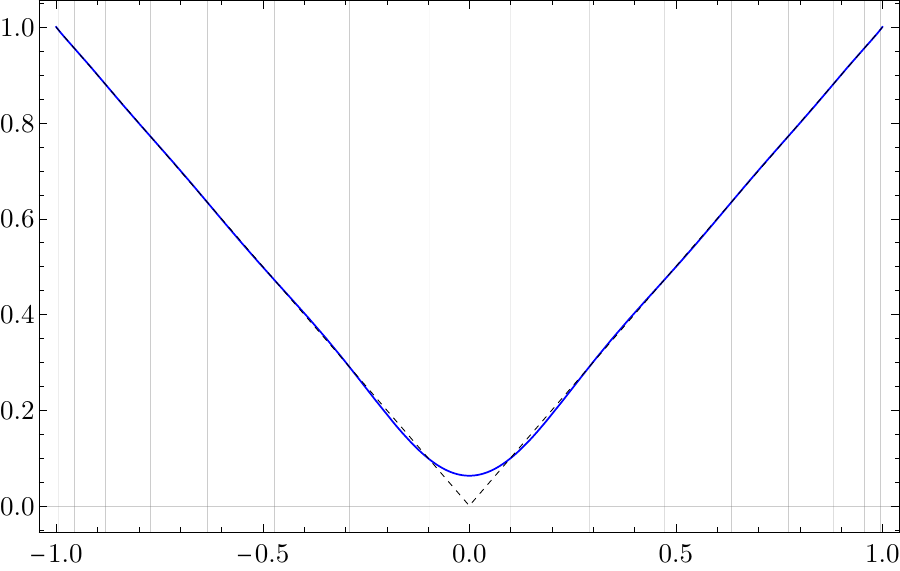}
	\caption{Interpolating polynomials of degree 15 for $f(x)=|x|$. The polynomials are shown as solid blue lines, while $f(x)$ is shown as dashed black lines. Thin vertical lines indicate the positions of the interpolation nodes. Left: the uniform choice of interpolation nodes~\eqref{eq:uniform}. Right: the Chebyshev nodes~\eqref{eq:chebyshev}.\label{fig:simpleinterpolation}}
\end{figure}

We illustrate this in figure~\ref{fig:simpleinterpolation}, where we compare the degree-15 interpolating polynomials for $f(x)=|x|$ for both choices of interpolation nodes. It is clear that the uniform choice~\eqref{eq:uniform} works very poorly. In fact, its interpolation error grows exponentially as $2^n$ with $n$ (times power-law factors), exceeding $10^{24}$ for $n=100$. On the other hand, the Chebyshev nodes~\eqref{eq:chebyshev} perform very well with an error that decays as $1/n$.\footnote{Both choices of interpolation nodes perform much better for smooth or analytic $f(x)$. We chose a non-differentiable function for illustration purposes. The Chebyshev nodes are generally exponentially better than the uniform nodes (except in degenerate cases such as when $f$ is a polynomial).}

We can quantify the difference between the choices~\eqref{eq:uniform} and~\eqref{eq:chebyshev} using the Lebesgue constant $\L_n$, which is defined as the operator norm of the interpolation operator $I_n$,
\be
	\L_n = \|I_n\|.
\ee
Recall that the operator norm is defined as
\begin{align}
	\|I_n\|=\sup_{\|f\|=1}\|I_n f\|.
\end{align}

An important property of the Lebesgue constant is that it bounds the interpolation error relative to the best approximating polynomial. Concretely, let $f\in C(U)$ and let $p_0(f)\in P_n$ be the polynomial which minimizes the error $\|p-f\|$ over $p\in P_n$. It is then easy to show that\footnote{Indeed, using $I_n p_0(f) = p_0(f)$ we find $\|I_n f - f\|\leq \|I_n f- p_0(f)\|+\|p_0(f)-f\|= \|I_n (f- p_0(f))\|+\|p_0(f)-f\|\leq (\L_n+1)\|p_0(f)-f\|.$\label{foot:proofLebesgue}}
\be\label{eq:LebesgueUpper}
	\|I_n f - f\| \leq (\L_n+1)\|p_0(f)-f\|, \quad \forall f\in C(U).
\ee
Conversely, for any given $n$ there exists a function $f\in C(U)$ for which\footnote{We restrict to $f$ for which $p_0(f)=0$. By the equioscillation theorem, these are precisely the functions $f$ such that $f(y_k)=\pm \|f\|$ for $n+2$ points $y_k\in [-1,1]$, with the sign alternating between the successive points. Let $g$ be a function such that $\|I_n g\|= \L_n\|g\|$ (it can be shown that the supremum in $\L_n=\|I_n\|$ is attained). Choosing $y_k$ disjoint from $x_i$, we can always modify $g$ in small neighbourhoods of $y_k$ to make sure $g(y_k)=\pm \|g\|$ as required by the equioscillation theorem. This does not modify $I_ng$ or $\|g\|$ and so $\|I_n g\|=\L_n\|g\|$ still holds. Setting $f=g$, we find $\|I_n f\|=\L_n\|f\|$ and $p_0(f)=0$. Using $\|I_n f-f\|\geq \|I_n f\|-\|f\|$ completes the proof.
}
\be\label{eq:LebesgueLower}
\|I_n f - f\| \geq (\L_n-1)\|p_0(f)-f\|.
\ee
Another useful property of $\L_n$ is related to the condition number of the map $p\mapsto (p(x_i))_{i=0}^n$; we will describe it later in this section.

It is possible to show (see \cite{BrutmanLebesgue} and references therein) that for the uniform nodes~\eqref{eq:uniform} and at large $n$
\be
	\L_n\sim \frac{2^{n+1}}{e n\log n},
\ee
while for the Chebyshev nodes~\eqref{eq:chebyshev}
\be\label{eq:LnCheb}
	\L_n \sim \frac{2}{\pi} \log(n+1).
\ee
Using~\eqref{eq:LebesgueUpper} and~\eqref{eq:LebesgueLower} we find that for large $n$ there exist functions $f$ for which the uniform nodes perform $\sim 2^n/(n\log^2 n)$ times worse than the Chebyshev nodes. 

We therefore see that it is desirable to minimize the Lebesgue constants $\L_n$, and in particular to get rid of the exponential growth that plagues the uniform interpolation nodes. As we will see below, the exponential growth of $\L_n$ only depends on the asymptotic density of nodes at large $n$. Denoting this density by $\r(x)$ and normalizing it to have total weight 1, we find $\r(x)=\frac{1}{2}$ for the uniform nodes~\eqref{eq:uniform} and 
\be\label{eq:chebdensity}
	\r(x)=\frac{1}{\pi\sqrt{1-x^2}}
\ee
for the Chebyshev nodes~\eqref{eq:chebyshev}. In fact, the density~\eqref{eq:chebdensity} is the unique density for which $\L_n$ do not grow exponentially. 

The key property of the Chebyshev nodes is therefore that they are distributed according to~\eqref{eq:chebdensity}. Any other choice of interpolation nodes with the density~\eqref{eq:chebdensity} would also work reasonably well for interpolation on $[-1,1]$ with uniform error norm. The Chebyshev nodes additionally cancel various subleading terms in $\L_n$, so that only the logarithmic growth~\eqref{eq:LnCheb} is present, and are generally the best practical choice in this situation. It should be noted, however, that they do not minimize $\L_n$, and the exact optimal interpolation nodes are not known in closed form. It is known that the optimal Lebesgue constants also satisfy~\eqref{eq:LnCheb}~\cite{Ber31}.

Our goal will therefore be to find the analogue of the density~\eqref{eq:chebdensity} for the interpolation problem on $U=[0,+\oo)$ with the error norm given in~\eqref{eq:bootstrapnorm}. In the remainder of this subsection we review how the Lebesgue constants can be computed and establish a useful property of $\L_n$. We will then analyze their large-$n$ behavior in the next subsection.

From now on, we consider general closed $U\subseteq \R$ and the norms of the form
\be
	\|f\| = \sup_{x\in U}\mu_n(x)|f(x)|,
\ee
where $\mu_n(x)>0$ is a fixed function that we allow to depend on the polynomial degree $n$. To compute the Lebesgue constant, we first define the interpolating basis functions
\be
	\ell_k(x) = \prod_{i\neq k}\frac{x-x_i}{x_k-x_i}.
\ee
The Lebesgue function is defined as
\be\label{eq:lebesguefn}
	\l_n(x)= \sum_{k=0}^n \frac{\mu_n(x)}{\mu_n(x_k)}|\ell_k(x)|.
\ee
The Lebesgue constant can then be obtained as
\be\label{eq:Lebesgue_constant}
	\L_n=\sup_{x\in U}\l_n(x).
\ee
Indeed, we have
\be
	\mu_n(x)|(I_nf)(x)|& = \mu_n(x)\left|\sum_{k=0}^n f(x_k)\ell_k(x)\right|\leq \sum_{k=0}^n \mu_n(x)|f(x_k)||\ell_k(x)|\nn\\
	&\leq \|f\|\sum_{k=0}^n\frac{\mu_n(x)}{\mu_n(x_k)}|\ell_k(x)|=\|f\|\l_n(x).
\ee
Taking $\sup_{x\in U}$ on both sides shows that $\|I_n f\|\leq \|f\|\sup_{x\in U}\l_n(x)$ and thus $\L_n\leq \sup_{x\in U}\l_n(x)$. Conversely, let $x_*$ be the position of the maximum of $\l_n(x)$ in $U$ and construct an $f\in C(U)$ such that $\mu(x_k)|f(x_k)|=\|f\|=1$ and $f(x_k)\ell_k(x_*)\geq 0$ for $k=0,\ldots,n$. We then have
\be
	\L_n\geq \|I_n f\|\geq \mu_n(x_*)|(I_nf)(x_*)|=\mu_n(x_*)\sum_{k=0}^n |f(x_k)||\ell_k(x_*)|=\l_n(x_*)=\sup_{x\in U}\l_n(x).
\ee
Together with $\L_n\leq \sup_{x\in U}\l_n(x)$ this proves $\L_n=\sup_{x\in U}\l_n(x)$.

Finally, the useful property of $\L_n$ alluded to above is related to the linear isomorphism
\begin{align}
	\f&:P_n\to \R^{n+1},\nn\\
	\f_k(p)&=\mu_n(x_k)p(x_k),
\end{align}
where $\R^{n+1}$ is equipped with the sup norm, $\|v\|=\max_k |v_k|$.
We find
\begin{align}\label{eq:phinorms}
	\|\f^{-1}\|=\L_n,\quad \|\f\|\leq 1.
\end{align}
Indeed,
\begin{align}
	\mu_n(x) p(x) = \mu_n(x)\sum_{k=0}^n \frac{\f_k(p)}{\mu_n(x_k)}\ell_k(x) 
	= \sum_{k=0}^n \f_k(p)\frac{\mu_n(x)}{\mu_n(x_k)}\ell_k(x)\leq \|\f(p)\|\l_n(x).
\end{align}
Therefore, $\|p\|\leq \L_n\|\f(p)\|$. On the other hand, the above inequality is saturated for some $x$ if we choose $\f_k(p)$ to be $k$-independent. This shows $\|\f^{-1}\|=\L_n$. For $\|\f\|$, the inequality $\|\f(p)\|\leq \|p\|$ is obvious since the supremum on both sides is taken of the same function, but on the left it is over a smaller set. 

We thus conclude that the condition number $\kappa(\f)=\|\f\|\|\f^{-1}\|$ of $\f$ satisfies
\begin{align}
	1\leq \kappa(\f)\leq \L_n.
\end{align}
This property will be useful in the discussion of section~\ref{sec:bilinearbasis}.

\subsection{Large-degree behavior of the Lebesgue constants}

Our goal now is to determine the conditions under which the Lebesgue constant $\L_n$ does not grow exponentially in the large-$n$ limit, assuming that the interpolation nodes $x_i$ are distributed according to a smooth density $\r(x)$. Since we will numerically test the interpolation nodes that follow from our analysis, we will not attempt to present a rigorous derivation. Instead, we will provide some basic intuition.

We expect that generically the maximum $\L_n$ of $\l_n(x)$ grows exponentially with $n$. At the same time, we know that $\l_n(x_i)=1$ for $i=0,\dots, n$, since all terms but one vanish in the sum in~\eqref{eq:lebesguefn}. This means that $\l_n(x)$ is oscillating. We are interested in finding the conditions under which the envelope of $\l_n(x)$ does not grow exponentially.

It is convenient to define the function $w_n(x) = \prod_{i=0}^n (x-x_i)$. In terms of $w_n(x)$, we have 
\be
	\ell_k(x) = \frac{w_n(x)}{(x-x_k)w'_n(x_k)}.
\ee
The Lebesgue function then becomes
\be\label{eq:ln_with_W}
	\l_n(x) = \mu_n(x)|w_n(x)|\sum_{k=0}^n\frac{1}{\mu_n(x_k)|x-x_k||w'_n(x_k)|}.
\ee
The local maxima of the function $\mu_n(x)|w_n(x)|$ between the interpolation nodes exhibit an exponential dependence on $n$, i.e. 
\be
	\mu_n(x)|w_n(x)|=e^{n F(x)}\x (\text{sub-exponential and oscillatory in $x$}).
\ee
The sum in~\eqref{eq:ln_with_W}, on the other hand, cannot produce such a dependence on $x$, and therefore we find for some constant $c$
\be\label{eq:lambdax_eq}
	\l_n(x) = e^{n (F(x)+c)}\x (\text{sub-exponential and oscillatory in $x$}).
\ee

We expect that for reasonable choices of $\mu_n$, $\L_n$ is minimized precisely when $\l_n(x)$ has equally large local maxima between all the consecutive pairs of interpolation nodes. The intuition is that if some maximum is larger than the others, we can move the two nodes adjacent to it closer to each other. This will decrease the value at this local maximum, even if at the cost of slightly increasing other maxima. Together with the requirement that $\L_n$ does not grow exponentially, this implies that $F(x)=-c$ on $[x_0,x_n]$.\footnote{In the case of the uniform sup norm on $U=[-1,1]$ it is known that for optimal nodes all the local maxima of $\l_n(x)$ are equal.} Outside of $[x_0,x_n]$ $F(x)$ can be non-constant, but $F(x)+c\leq 0$ has to be satisfied.

It is easy to obtain an explicit expression for $F(x)$. We have
\be
	\log \mu_n(x)+\sum_{i=0}^n \log|x-x_i|=n F(x) +\ldots.
\ee
Replacing the sum by an integral and defining $W_n(x) = -2\log \mu_n(x)$, we obtain
\be\label{eq:Fxeq}
	F(x) =-\frac{1}{2(n+1)}W_n(x)+\int dx' \rho(x')\log|x-x'|,
\ee
where $\r(x)$ is the density of interpolation nodes. Note that we used $n+1$ instead of $n$ in the denominator for future convenience; both are equivalent at this order. Requiring $F'(x)=0$ is now equivalent to
\be\label{eq:samplepointseq}
	\frac{1}{2(n+1)}W'_n(x) = \mathrm{p.v.}\int \frac{\rho(x')dx'}{x-x'}.
\ee
In the next subsection we will solve this equation for $\rho(x')$ when $\mu_n(x)$ is given by~\eqref{eq:muforconformalblocks}.

\subsection{Optimal interpolation nodes for the numerical bootstrap}
\label{sec:optimalnodes}

We now explain how to determine the density $\r(x)$. First, note that \eqref{eq:samplepointseq} has a nice physical interpretation: It describes the equilibrium density of charges with logarithmic pairwise interactions in an external potential $\frac{1}{2(n+1)}W_n(x)$. Because of this, it has already been studied in many different physical contexts. For example, it appears when solving for the density of eigenvalues in matrix models \cite{Marino:2004eq}. See also \cite{saff2013logarithmic} for a careful analysis using potential theory, and \cite{muskhelishvili1953singular} for other applications in physics. Here, we will mainly focus on describing the solution to \eqref{eq:samplepointseq}, and we refer the reader to the aforementioned references for a more rigorous derivation.

The main idea is to consider a function $\omega(x)$ defined as\footnote{In the matrix model language, $\omega(x)$ is the genus zero resolvent.}
\be\label{eq:omega_def}
	\omega(x) = \int \frac{\r(x')dx'}{x-x'}.
\ee
Using the identity $\frac{1}{x+i\e} =\mathrm{p.v.} \frac{1}{x}-i\pi\de(x)$, the density $\r(x)$ can be written as
\be
	\r(x) = -\frac{1}{2\pi i}\p{\omega(x+i\e) - \omega(x-i\e)}.
\ee
In terms of $\omega(x)$, \eqref{eq:samplepointseq} becomes
\be\label{eq:samplepointeq_inomega}
	\frac{1}{2(n+1)}W'_n(x) = \frac{1}{2}\p{\omega(x+i\e) + \omega(x-i\e)}.
\ee

For our choice of function $\mu_n(x)$, it turns out that the solution $\r(x)$ is supported on a closed interval $\mathrm{supp}(\r) = [a,b]$, and one should solve \eqref{eq:samplepointeq_inomega} only for $x$ in this interval. Moreover, the corresponding $\omega(x)$ can be written as an integral over this closed interval. More precisely, the solution to \eqref{eq:samplepointeq_inomega} is given by
\be\label{eq:omega_solution}
	\omega(x) = \frac{1}{n+1}\int_a^b \frac{dx'}{2\pi} \frac{W'_{n}(x')}{x-x'}\frac{((x-a)(x-b))^{\frac{1}{2}}}{((x'-a)(b-x'))^{\frac{1}{2}}}.
\ee
The interval $[a,b]$ can be determined by studying the behavior of $\omega(x)$ in the $x\to \oo$ limit. From the definition \eqref{eq:omega_def}, we have $\omega(x) = \frac{1}{x} + O(\tfrac{1}{x^2})$ in $x \to \oo$. Therefore, taking the $x\to \oo$ limit of \eqref{eq:omega_solution}, we obtain
\be
	 0=&\frac{1}{n+1}\int_a^b \frac{dx'}{2\pi}\frac{W'_{n}(x')}{((x'-a)(b-x'))^{\frac{1}{2}}} , \label{eq:intervaleq1} \\
	 1=&\frac{1}{n+1}\int_a^b \frac{dx'}{2\pi}\frac{x' W'_{n}(x')}{((x'-a)(b-x'))^{\frac{1}{2}}} . \label{eq:intervaleq2}
\ee

It now seems straightforward to find the density $\r(x)$. We should plug our $W_n(x)$ into \eqref{eq:intervaleq1} and \eqref{eq:intervaleq2} to solve for $a$ and $b$, compute $\omega(x)$ using \eqref{eq:omega_solution}, and finally take the discontinuity to get $\r(x)$. However, there are still some subtleties we have to be careful about. Note that our allowed range of $x$ is $U=[0,\oo)$, and so we can imagine the potential $W_n(x)$ is infinite for negative $x$. On the other hand, we also expect that the left end of the interval $[a,b]$ is $a=0$, which is the boundary of $U$. In this case, there can be additional boundary term contributions to \eqref{eq:omega_solution}, \eqref{eq:intervaleq1}, and \eqref{eq:intervaleq2}.

To fix these boundary terms, let us introduce a small regulator $\e$ and consider $\mu^{(\e)}_n(x) \equiv x^{(n+1) \e} \mu_n(x)$. This gives $W^{(\e)}_n(x) = -2(n+1)\e \log x + W_n(x)$. For any $\e>0$, the regulated potential $W^{(\e)}_n(x)$ goes to infinity when we approach $x=0$. Therefore the regulated equilibrium interval $[a^{(\e)},b^{(\e)}]$ never touches the boundary of $U$ (i.e., $a^{(\e)}>0$), and we can apply \eqref{eq:omega_solution}, \eqref{eq:intervaleq1}, and \eqref{eq:intervaleq2}. The boundary terms in the unregulated case then come from carefully taking the $\e\to 0$ limit. Plugging the regulated potential into \eqref{eq:intervaleq1} and \eqref{eq:intervaleq2}, we have
\be
	0=&-2\int_{a^{(\e)}}^{b^{(\e)}}\frac{dx'}{2\pi}\frac{\e}{x'}\frac{1}{\p{(x'-a^{(\e)})(b^{(\e)}-x')}^{\frac{1}{2}}} + \frac{1}{n+1}\int_{a^{(\e)}}^{b^{(\e)}}\frac{dx'}{2\pi}\frac{W'_n(x')}{\p{(x'-a^{(\e)})(b^{(\e)}-x')}^{\frac{1}{2}}}, \label{eq:interval_eq1_witheps} \\
	1=&-2\int_{a^{(\e)}}^{b^{(\e)}}\frac{dx'}{2\pi}\frac{\e}{\p{(x'-a^{(\e)})(b^{(\e)}-x')}^{\frac{1}{2}}} + \frac{1}{n+1}\int_{a^{(\e)}}^{b^{(\e)}}\frac{dx'}{2\pi}\frac{x' W'_n(x')}{\p{(x'-a^{(\e)})(b^{(\e)}-x')}^{\frac{1}{2}}}. \label{eq:interval_eq2_witheps}
\ee
In the $\e \to 0$ limit, we find
\be
	a^{(\e)} = \frac{\e^2}{b R(b)^2} + O(\e^3), \qquad b^{(\e)} = b+ O(\e),
\ee
where $b$ and $R(b)$ depend on details of $W_n(x)$.\footnote{For example, for $\mu_n(x)=r^x$, the interval has a closed-form expression given by
\be
	[a^{(\e)},b^{(\e)}] = \left[\tfrac{n+1}{-\log r}\p{1+\e-\sqrt{1+2\e}},\tfrac{n+1}{-\log r}\p{1+\e+\sqrt{1+2\e}}\right],
\ee
which becomes $a^{(\e)} = \frac{n+1}{-2\log r}\e^2 + O(\e^3)$ and $b^{(\e)} = \frac{2(n+1)}{-\log r} +O(\e)$ in the $\e\to 0$ limit.
} Let us focus on the first term on the right-hand side of \eqref{eq:interval_eq1_witheps}. After making a change of variable $x=y^2+a^{(\e)}$ and taking the $\e\to 0$ limit, it becomes
\be
	- \lim_{\e\to 0}\int_{-\sqrt{b^{(\e)}-a^{(\e)}}}^{\sqrt{b^{(\e)}-a^{(\e)}}}\frac{dy}{\pi}\frac{\e}{y^2+a^{(\e)}}\frac{1}{(b^{(\e)}-a^{(\e)}-y^2)^{\frac{1}{2}}} = - R(b)\int dy\frac{ \de(y)\sqrt{b}}{(b-y^2)^{\frac{1}{2}}} = -R(b).
\ee
We see that even after taking $\e \to 0$ limit, the regulator $x^{n\e}$ has a nonvanishing contribution $-R(b)$. Alternatively, one can imagine that the potential derivative $W'_n(x)$ has a delta-function term $-(n+1)\pi \sqrt{b} R(b)\de(\sqrt{x})$. A similar argument shows that $\omega(x)$ given by \eqref{eq:omega_solution} also receives a nonvanishing contribution $-R(b)\sqrt{\frac{x-b}{x}}$ in the $\e\to 0$ limit.

In summary, when we restrict $x$ to be positive, $\omega(x)$ should be given by
\be\label{eq:omega_solution_withRb}
	\omega(x) = -R(b)\sqrt{\frac{x-b}{x}} + \frac{1}{n+1}\int_0^b \frac{dx'}{2\pi} \frac{W'_{n}(x')}{x-x'}\frac{(x(x-b))^{\frac{1}{2}}}{(x'(b-x'))^{\frac{1}{2}}},
\ee
and the interval $[0,b]$ and $R(b)$ can be determined using
\be
	0=& -R(b)+\frac{1}{n+1}\int_0^b \frac{dx'}{2\pi}\frac{W'_{n}(x')}{(x'(b-x'))^{\frac{1}{2}}},  \\
	 1=& \frac{1}{n+1}\int_0^b \frac{dx'}{2\pi}\frac{x' W'_{n}(x')}{(x'(b-x'))^{\frac{1}{2}}}.
\ee

To obtain a final answer for the density $\r(x)$, we must plug in an actual expression for $W_n(x)$. Let us consider an ansatz
\be
\label{eq:ansatzformu}
\mu_n(x) &= \frac{r^x}{\prod_i (x+k_i)^{\a_i}}, \quad\textrm{where}\quad r,\a_i \in \R\textrm{, and } k_i \in \R_{\geq 0}.
\ee
This ansatz includes in particular the case (\ref{eq:muforconformalblocks}) relevant for approximating conformal blocks. We find
\be
	W_n(x) = 	-2 \log r\ x + 2\sum_i \a_i \log(x+k_i).
\ee
This gives a simple equation for $b$,
\be\label{eq:b_equation}
	-\frac{b}{2}\log r + \sum_i\a_i\p{1-\sqrt{\frac{k_i}{b+k_i}}}=n+1,
\ee
and $R(b)$ is given by
\be
	R(b) = -\frac{\log r}{n+1} + \frac{1}{n+1}\sum_i \frac{\a_i }{\sqrt{k_i(b+k_i)}}.
\ee
Taking the discontinuity of \eqref{eq:omega_solution_withRb}, we find
\be\label{eq:rhox_solution}
	\r(x) = \frac{1}{(n+1)\pi}\sqrt{\frac{b-x}{x}}\p{-\log r + \sum_i \a_i\sqrt{\frac{k_i}{b+k_i}}\frac{1}{x+k_i}}.
\ee

Finally, we would like to turn the density $\r(x)$ into a discrete set of interpolation nodes. We do this by imposing a condition similar to the Bohr-Sommerfeld quantization condition. Let us consider the function $\cN(x)\equiv (n+1)\int_0^{x}\r(x')dx'$. It is given by
\be\label{eq:N_expression}
	\cN(x) = &\frac{-\log r}{\pi}\p{\sqrt{x(b-x)}+\frac{b}{2}\cos^{-1}\p{1-\frac{2x}{b}}}\nn \\
	&+\sum_i\frac{\a_i}{\pi}\p{\cos^{-1}\p{1-\frac{2x(k_i+b)}{b(k_i+x)}}-\sqrt{\frac{k_i}{b+k_i}}\cos^{-1}\p{1-\frac{2x}{b}}}.
\ee
In the discrete case, $\cN(x)$ is a sum of step functions $\sum_i \th(x-x_i)$, where $x_i$'s are the locations of the interpolation nodes. The above expression of $\cN(x)$ gives a smooth approximation for the discrete sum, and it should be a half-integer at the location of each $x_i$. In other words, we obtain the interpolation nodes $x_i$'s using the following quantization condition:
\be\label{eq:quantization_condition}
	\cN(x_i) = i+\frac{1}{2},\qquad i=0,\ldots,n.
\ee
In what follows, the interpolation nodes obtained in this way will be referred to as the optimal density interpolation nodes.

\begin{figure}
	\centering
	\includegraphics[scale=0.8]{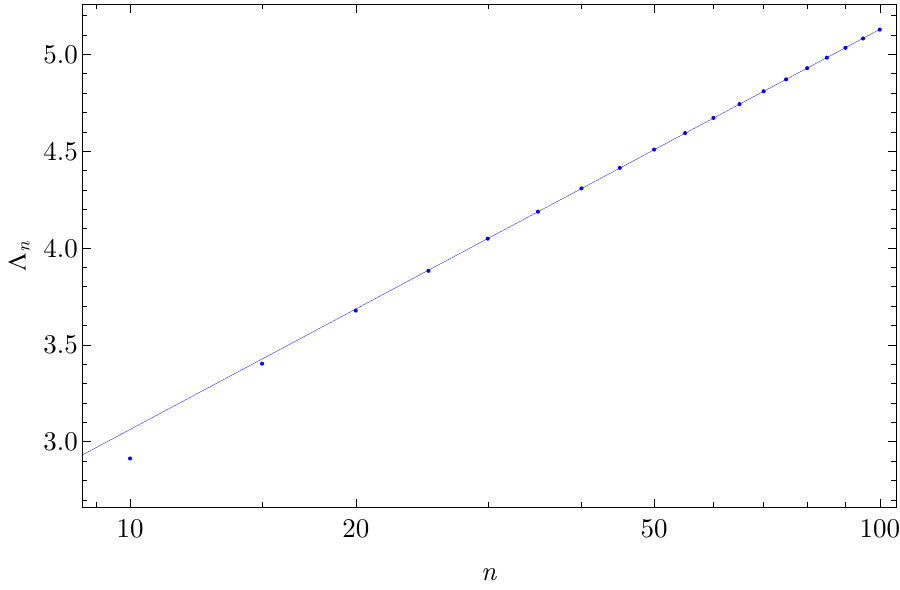}
	\caption{Lebesgue constants for $\mu_n(x)=r^x/\prod_{k=1}^m(x+k)$, where $n=m+10$, in log-linear scale. The line is a fit of the form $a\log n+b$. In the fit $a=0.897, b=0.997$. For comparison, for Laguerre-like sample points previously used by \texttt{SDPB} (not for block approximation, see section~\ref{sec:internals}) $\L_n=7.6\x 10^{24}$ at $n=30$.}
	\label{fig:LebesguePlot_samplepoints}
\end{figure}

Let us consider a concrete example to illustrate how the optimal density interpolation nodes can improve the behavior of the Lebesgue constant. We consider the case
\be
	\mu_n(x) = \frac{r^x}{\prod_{k=1}^m(x+k)},\quad n=m+10.
\ee
For each $n$, we use \eqref{eq:b_equation}, \eqref{eq:N_expression}, and \eqref{eq:quantization_condition} to determine the interpolation nodes $x_0,\ldots,x_{n}$. The Lebesgue constant $\Lambda_n$ is then computed using \eqref{eq:lebesguefn} and \eqref{eq:Lebesgue_constant}. The result is shown in figure \ref{fig:LebesguePlot_samplepoints}. We see that the Lebesgue constant for the optimal density interpolation nodes grows only logarithmically at large $n$. 

\subsection{Tests of optimal interpolation nodes}

As explained in section~\ref{sec:errornorm}, we seek approximations to conformal blocks of the form
\be
4^{-\De} \left.\ptl_z^m\ptl_{\bar z}^n g_{\De,j}(z,\bar z)\right|_{z=\bar z = \frac 1 2} &\sim r_*^\De \frac{P^{m,n}_j(\De)}{\prod_{i=1}^{n_\mathrm{poles}} (\De-\De_i)}.
\label{eq:thingwewanttodoagain}
\ee
We will choose the denominator to be a product of the $n_\mathrm{poles}$ largest poles (i.e.\ farthest to the right on the number line) of the conformal block.  We define $\mu^{(j)}(\De)$ as in (\ref{eq:muforconformalblocks}). This has the form (\ref{eq:ansatzformu}) (where $\De=\De_\mathrm{min}(j)+x$), so we can compute the optimal density of interpolation nodes using (\ref{eq:rhox_solution}), and the locations of the interpolation nodes using~(\ref{eq:quantization_condition}).

\subsubsection{Conformal blocks on the diagonal}

Here we illustrate the above method as applied to computing improved approximations to 3d conformal blocks with identical external scalars. In radial coordinates $(r,\eta)$ such blocks can be written along the diagonal as~\cite{Rychkov:2015lca}
\begin{align}
g^{(d=3)}_{\De,j}(r, \eta=1) &= \frac{(4r)^{\De}}{1-r^2} {}_4F_3\left(\left\{\frac12, \frac{\De-j-1}{2},\frac{\De+j}{2},\De-1\right\},\left\{\De - \frac12, \frac{\De+j+1}{2}, \frac{\De-j}{2}\right\}, r^2 \right).
\label{eq:3dblockdiagonal}
\end{align}
More generally 3d conformal blocks can be computed using the recursion relations derived in~\cite{Kos:2013tga, Kos:2014bka, Penedones:2015aga, Erramilli:2019njx}.

Expanding~(\ref{eq:3dblockdiagonal}) in the $r$-expansion produces a sum over poles in $\De$, which at a fixed order get resummed to the form $g^{(d=3)}_{\De,j} \approx (4r)^\De \frac{P(\De)}{\prod_i (\De - \De_i)}$ for some polynomial numerator $P(\De)$. This approximation can be made arbitrarily precise at the expense of increasing the order of $P(\De)$. We can then reduce the order of the polynomial numerator using interpolation: we write a new ansatz by taking the $n_{\text{poles}}$ rightmost poles, $\tilde{g}^{(d=3)}_{\De,j} = (4r)^{\De} \frac{\tilde{P}(\De)}{\prod^{n_{\text{poles}}}_i (\De - \De_i)}$, and compute the coefficients in $\tilde{P}(\De) = \sum_n c_n \De^n$ via interpolation with the optimal density nodes for the weight $\mu^{(j)}(\De)$.

The result of this procedure can be compared to the older method of reducing the polynomial order (sometimes called ``pole-shifting") introduced in~\cite{Kos:2013tga}. In this method, one computes the new polynomial numerator by matching $\Delta$-derivatives, half computed by matching at the unitarity bound and half at $\Delta \rightarrow \infty$.\footnote{When there are an odd number of coefficients one includes an extra constraint at the unitarity bound.} Pole shifting gives very good conformal block approximations at these endpoints, but allows larger errors at intermediate $\Delta$.

\begin{figure}[t]
	\centering
	\includegraphics[width=0.49\textwidth]{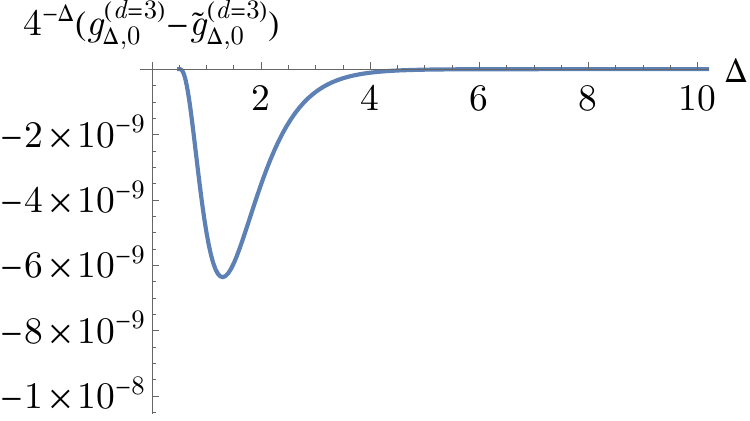}
	\includegraphics[width=0.49\textwidth]{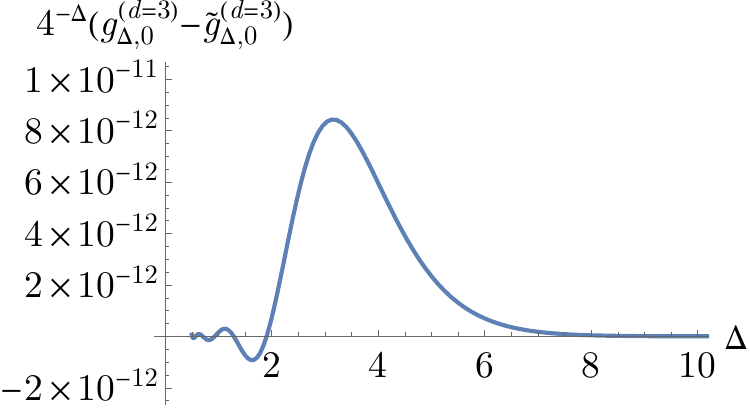}
	\caption{Approximation error $4^{-\De}(g_{\De,j} - \tilde{g}_{\De,j})$ as a function of $\Delta$ for the 3d conformal block with $j=0$, approximated using 6 poles in the crossing symmetric configuration $(r,\eta) = (r_*,1)$. We compare the approximation error using the older pole-shifting method (left), which matches derivatives of the block at $\De = 1/2, \infty$, against the new method which matches at the optimal interpolation nodes (right).\label{fig:3dblockerror6poles}}
\end{figure}

\begin{figure}[t]
	\centering
	\includegraphics[width=\textwidth]{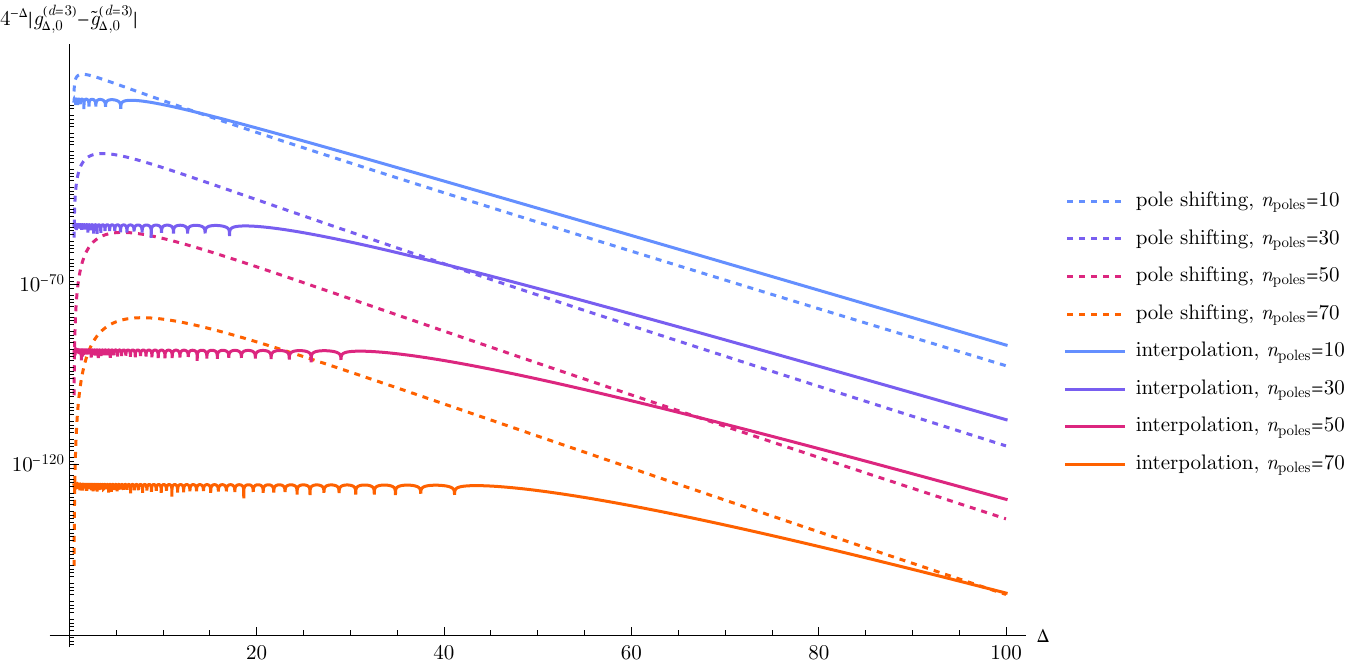}
	\caption{Approximation error $4^{-\De}|g_{\De,j} - \tilde{g}_{\De,j}|$ as a function of $\Delta$ for the 3d conformal block with $j=0$, approximated using $n_\mathrm{poles}=10,30,50,70$, using both pole-shifting (first four curves) and interpolation with optimal density nodes (last four curves).\label{fig:3dblockerrormorepoles}}
\end{figure}

\begin{figure}[t]
	\centering
	\includegraphics[width=0.49\textwidth]{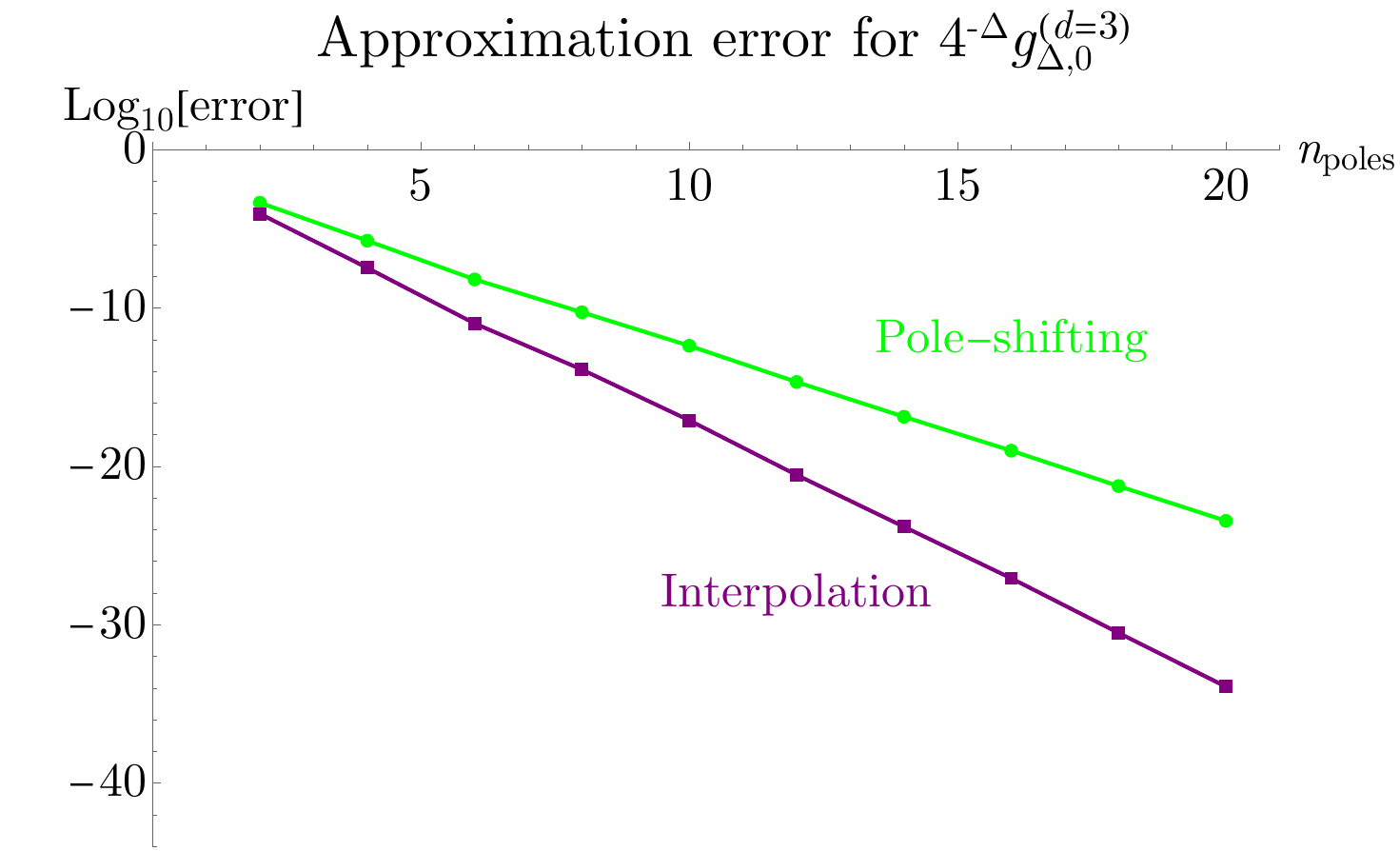}
	\includegraphics[width=0.49\textwidth]{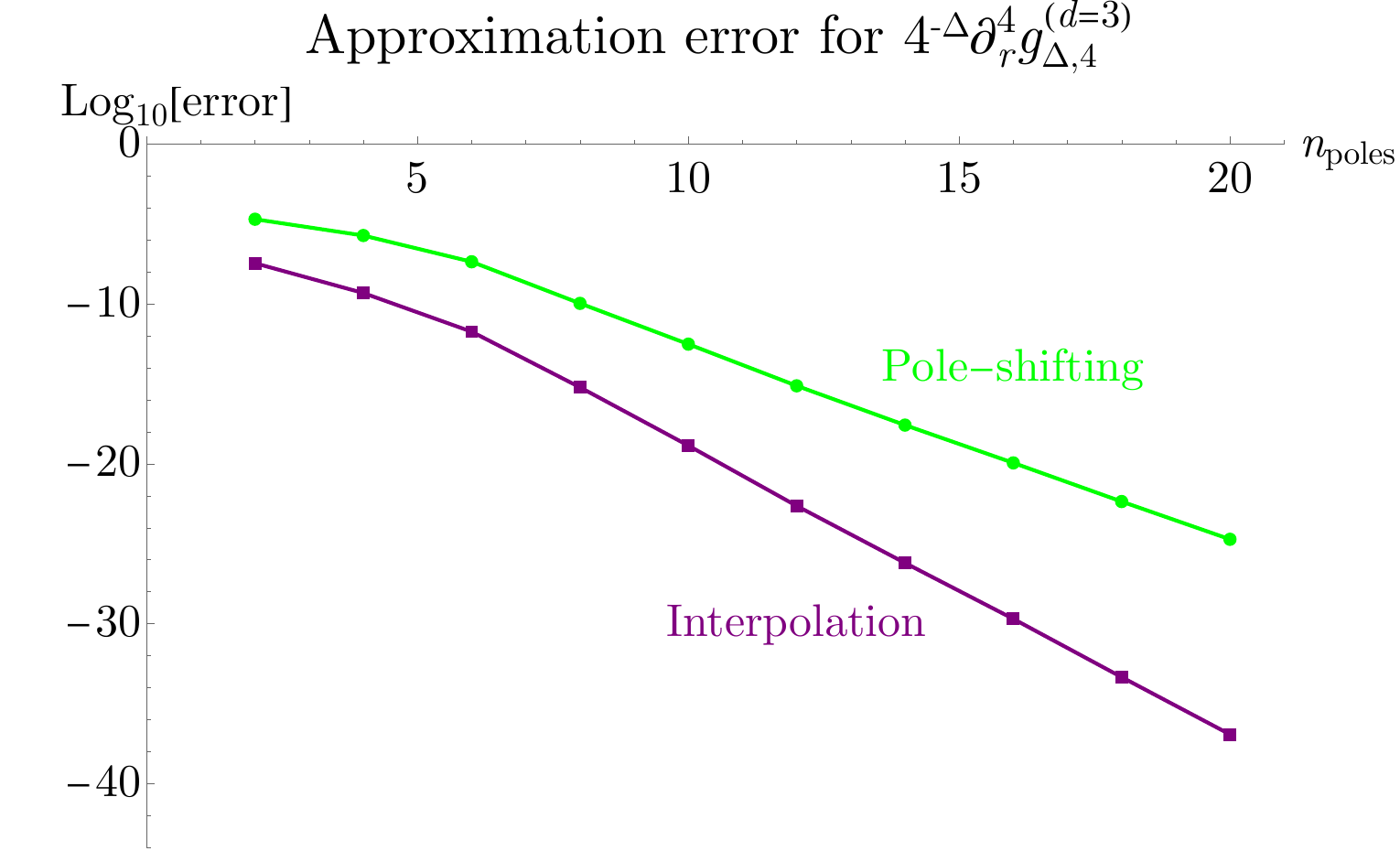}
	\caption{Log${}_{10}$ of the magnitude of the maximum approximation error $4^{-\De}|g^{(d=3)}_{\De,j} - \tilde{g}^{(d=3)}_{\De,j}|$ across $\Delta$ in the 3d conformal block with $j=0$ (left) and 4-derivatives of the block with $j=4$ (right), evaluated in the crossing symmetric configuration $(r,\eta) = (3-2\sqrt{2},1)$. We compare the approximation error using the optimal interpolation nodes (purple) with $n_{\text{poles}}$ kept poles against the older pole-shifting method (green) with the same number of poles.\label{fig:3dblockerror}}
\end{figure}

To give an example, let us consider approximating the $j=0$ block at the crossing-symmetric point $r=r_*$ using 6 poles, giving an approximation of the form\footnote{Here, we only include the poles with nonvanishing residues for identical external dimensions. For generic external dimensions, one should also include poles at $\De=0,-2,\ldots$. This choice of poles applies to all our approximations of the 3d block shown in figures~\ref{fig:3dblockerror6poles}, \ref{fig:3dblockerrormorepoles}, and \ref{fig:3dblockerror}, since \eqref{eq:3dblockdiagonal} does not contain these poles. On the other hand, when approximating the functional, we choose poles based on the procedure described in section \ref{sec:implementation}, which includes poles with vanishing residues for identical external dimensions (see also footnote \ref{note:polechoice}).} 
\begin{align}
\tilde{g}_{\De,0}^{(d=3)} &= (4r_*)^{\De} \frac{\sum_{n=0}^{6} c_n \De^n}{(\De-1/2)(\De+1/2)(\De+1)(\De+3/2)(\De+5/2)(\De+3)},
\end{align} 
for some coefficients $c_n$ that we can fix by matching to the exact answer using either derivatives at $\Delta = 1/2,\infty$ or by matching at the optimal interpolation nodes. In figure~\ref{fig:3dblockerror6poles} we show the error of each of these approximations as a function of $\De$. We see that, in this simple example, the maximum approximation error using the interpolation nodes is smaller by two orders of magnitude than the older pole-shifting approach.

In figure~\ref{fig:3dblockerrormorepoles}, we show a comparison between $g_{\De,0}^{(d=3)}$ and its approximation $\tl g_{\De,0}^{(d=3)}$ using larger numbers of poles: $n_\mathrm{poles}=10,30,50,70$, for both pole-shifting and interpolation.   Interpolation gives errors that are small and approximately constant over some range of $\De$, before exponentially decaying. Furthermore, interpolation improves faster than pole-shifting as $n_\mathrm{poles}$ is increased. The tradeoff is that interpolation is somewhat less accurate at very small and large $\De$. However, we will confirm with numerical tests in section~\ref{sec:numericaltests} that this tradeoff is worth making when computing bootstrap bounds.

In figure~\ref{fig:3dblockerror} we show additional comparisons between these two methods as $n_{\text{poles}}$ is increased. We plot the log of the magnitude of the maximum approximation error $4^{-\De}|g^{(d=3)}_{\De,j} - \tilde{g}^{(d=3)}_{\De,j}|$ across $\De$ in the crossing symmetric configuration. As illustrative examples we show the scalar block $j = 0$ and 4 $r$-derivatives of the $j = 4$ block, but other blocks and choices of derivatives show similar behavior. We can see clearly that matching at the optimal density interpolation nodes produces blocks with a maximum error that is smaller by several orders of magnitude compared to the pole-shifting method, with a faster exponential convergence as $n_{\text{poles}}$ is increased.

\subsubsection{The spin-2 gap optimal functional}

\begin{figure}[t]
	\centering
	\includegraphics[width=0.85\textwidth]{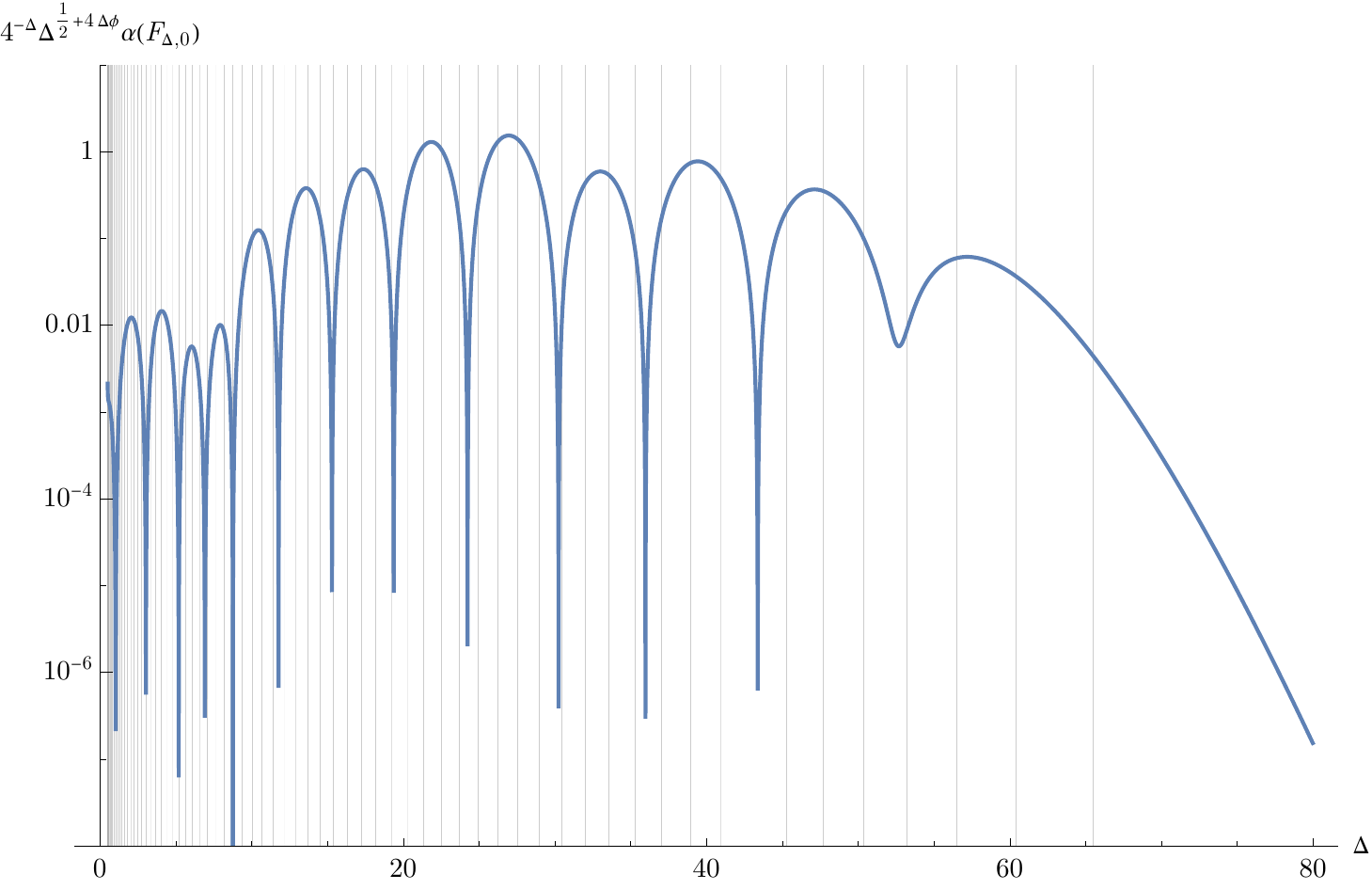}
	\caption{\label{fig:functionalwithsamplepoints}A plot of the normalized spin-2 gap functional with $\Lambda=59$, acting on blocks with spin $j=0$, with optimal density interpolation nodes for $n_\mathrm{poles}=10$ superimposed as grey lines.}
\end{figure}

Let us also revisit the spin-2 gap optimal functional at $\Lambda=59$ discussed in section~\ref{sec:errornorm}. In figure~\ref{fig:functionalwithsamplepoints}, we show a plot of the functional with the optimal density interpolation nodes for $n_\mathrm{poles}=10$ superimposed as grey vertical lines. As expected, the interpolation nodes are denser for smaller $\Delta$, where the functional has the most structure. They get sparser and sparser at large $\De$, and the rightmost sample point is at $\De\approx 65$, where the functional's oscillations stop and exponential decay begins.

\begin{figure}[t]
	\centering
	\includegraphics[width=\textwidth]{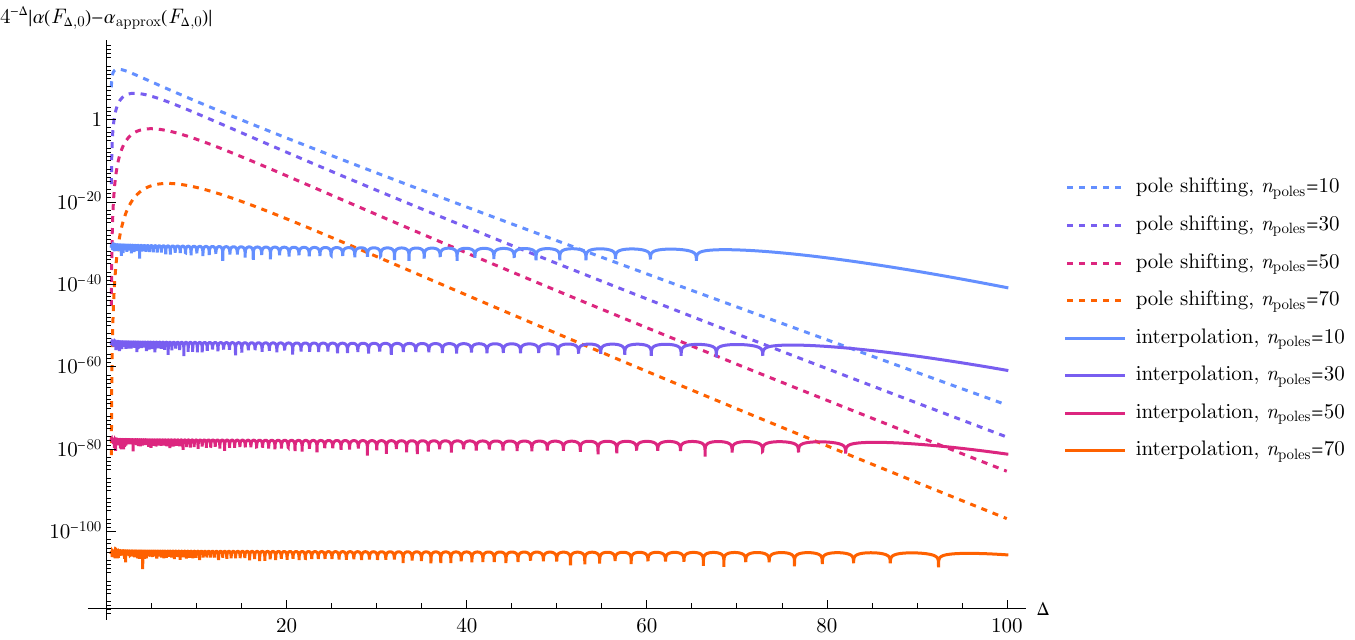}
	\caption{\label{fig:approximationcomparisons}A comparison of errors in $4^{-\De}\a(F_{\De,0})$ for different approximation schemes. The first four curves show pole-shifting with $n_\mathrm{poles}=10,30,50,70$, while the other four curves show interpolation (with optimal density nodes) with the same values of $n_\mathrm{poles}$.}
\end{figure}

In figure~\ref{fig:approximationcomparisons}, we show the difference between the approximated functional and the true functional, using both the old pole-shifting scheme and our new interpolation scheme, for different values of $n_\mathrm{poles}$. We again see that pole-shifting gives good approximations for small and large $\De$, but is inferior to interpolation for intermediate $\De$. Interpolation with optimal density nodes leads to errors that are nearly flat as a function of $\De$ in the regime where the functional has interesting structure (despite the interpolation nodes becoming sparser and sparser as we move to the right), and decay exponentially (like the functional itself) outside of this region. Furthermore, the errors decrease quickly with $n_\mathrm{poles}$. For example, interpolation with $n_\mathrm{poles}=70$ is sufficient to obtain uniform errors less than $10^{-105}$, while pole-shifting with $n_\mathrm{poles}=70$ leads to errors of size $10^{-17}$ at the worst value of $\De$.

\section{Condition numbers and implementation}
\label{sec:conditionnumbers}

In this section we discuss how interpolation nodes and other choices affect the condition numbers of the matrices that~\texttt{SDPB} operates on internally. Reducing these condition numbers is important for lowering precision requirements and therefore reducing the time it takes~\texttt{SDPB} to perform one iteration.

Empirically, we found that the worst condition number in a typical~\texttt{SDPB} run is that of the Schur complement matrix $S$~\cite{Simmons-Duffin:2015qma}. In subsection~\ref{sec:internals} we review the necessary facts about the internal logic of~\texttt{SDPB} and define the matrix $S$. In subsections~\ref{sec:bilinearbasis} and~\ref{sec:implementation} we describe the choices of the sample points, the bilinear bases, and the scaling factors, as implemented in~\texttt{SDPB 3.1} (in~\texttt{pmp2sdp}, to be more precise) as the result of this work.

\subsection{SDPB internals}
\label{sec:internals}

Let us first review some definitions related to the algorithm used in~\texttt{SDPB 3.0}, i.e.\ prior to this work. For a more detailed discussion see~\cite{Simmons-Duffin:2015qma,Landry:2019qug}.

The problem solved by~\texttt{SDPB} is a polynomial matrix program (PMP) of the following form,
\be
&\text{maximize  }b\.y\text{  over  }y\in \R^N\nn\\
&\text{such that  }M_j^0(x)+\sum_{k=1}^N y_k M^k_j(x)\succeq 0\text{  for all  }x>0\text{  and  }1\leq j\leq J.
\label{eq:PMP}
\ee
Here, the matrices $M_j^k(x)$ are symmetric matrices with polynomial entries, 
\be
	M_j^k(x) = \begin{pmatrix}
		P^k_{j,11}(x) & \cdots & P^k_{j,1m_j}(x)\\
		\vdots & \ddots & \vdots\\
		P^k_{j,m_j 1}(x) & \cdots & P^k_{j,m_jm_j}(x)\\
	\end{pmatrix},
\ee
where $P^k_{j,rs}(x)$ are polynomials. In numerical bootstrap applications $M_j^k(x)$ are polynomial approximations to conformal blocks. More precisely, one usually has
\be
	G_j^k(x)\approx \frac{cr^x}{\prod_i(x-p_i)^{\a_i}}M_j^k(x),
\ee
where $G_j^k(x)$ is a matrix of that consists of derivatives of conformal blocks with $\De=x+\text{const}$, $r=3-2\sqrt{2}$, $c=1$, and $p_i\leq 0$ is a set of poles with multiplicities $\a_i$. See, for example,~\cite{Kos:2013tga,Kos:2014bka} for details. As input \texttt{SDPB} (specifically, the \texttt{pmp2sdp} program) accepts not just the polynomial matrices $M_j^k(x)$ but also the prefactor
\begin{align}\label{eq:sdpbprefactor}
	\mu^{(j)}(x)=\frac{c_jr_j^x}{\prod_i(x-p_{j,i})^{\a_{j,i}}},
\end{align}
with $c_{j}, r_j, p_{j,i}, \a_{j,i}$ fully controlled by the user. 

This PMP is converted to a semidefinite program (SDP) of the form
\be
	&\text{maximize  }b\.y\text{  over  }y\in \R^N,\, Y\in  \mathcal{S}^K\nn\\
	&\text{such that  }\Tr\p{A_*Y}+By=c,\text{  and  }Y\succeq 0.
	\label{eq:dualSDP}
\ee
Here $c\in \R^P,\,B\in \R^{P\x N}, A_1,\ldots, A_P\in \mathcal{S}^K$, and $\mathcal{S}^K$ is the set of all symmetric $K\x K$ matrices. Furthermore, $\Tr\p{A_*Y}$ denotes the vector $(\Tr\p{A_1Y},\ldots, \Tr\p{A_PY})$. We refer to this SDP as the ``dual SDP".

The equivalence between the PMP~\eqref{eq:PMP} and the dual SDP~\eqref{eq:dualSDP} is explained in detail in~\cite{Simmons-Duffin:2015qma}.
Here we merely state the relationship between $A_p$ and the matrices $M_j^k$. It is fixed once a ``bilinear basis'' of polynomials $q_m^{(j)}(x)$, sample points $x_k^{(j)}$, and sample scalings $s_k^{(j)}$ are provided by the user.
The polynomials $q_m^{(j)}(x)$ must be defined for $m\geq 0$ and satisfy $\deg q_m^{(j)} = m$.

For every integer $\de>0$ we define $(\de+1)\x(\de+1)$ matrices,
\be\label{eq:Qdefn}
	(Q_\de^{(j)}(x))_{mn}\equiv q^{(j)}_m(x)q^{(j)}_n(x),\quad 0\leq m,n\leq \de,
\ee
where $1\leq j\leq J$. Furthermore, we define
\be
	(E^{rs})_{ij}=\thalf\p{\de^r_i\de^s_j+\de^r_j\de^s_i}.
\ee
Then we require
\be
	P^0_{j,rs}(x)+\sum_{k=1}^N y_k P^k_{j,rs}(x)=
	\Tr\p{Y_{2j-1}(Q_{\de_{j1}}^{(j)}(x)\otimes E^{rs})}
	+
	\Tr\p{Y_{2j}(xQ_{\de_{j2}}^{(j)}(x)\otimes E^{rs})}.
	\label{eq:Yyrelation}
\ee
Here $Y_l\in \mathcal{S}^K$ are the diagonal blocks of $Y$, i.e.\ $Y=Y_1\oplus \cdots \oplus Y_{2J}$. The polynomial degrees $\de_{j1}$ and $\de_{j2}$ are defined as
\be
\de_{j1}=\lfloor\tfrac{d_j}{2}\rfloor,\quad \de_{j2}=\lfloor\tfrac{d_j-1}{2}\rfloor,\quad d_j = \max_{k,r,s} \p{\deg P^k_{j,rs}(x)}.
\ee
Positive-semidefiniteness of $Y$ now guarantees that the inequality in~\eqref{eq:PMP} is satisfied.

The relation~\eqref{eq:Yyrelation} is an inhomogeneous linear constraint between $y$ and $Y$. Similarly, the requirement of $Y$ being block-diagonal is a linear constraint on $Y$. Therefore, all these conditions can be expressed in terms of the constraints that are allowed in the formulation of the SDP~\eqref{eq:dualSDP}, for some appropriate choices of $A_p, B, c$. In practice, only the relation~\eqref{eq:Yyrelation} is encoded in this way in~\texttt{SDPB}, while the block-diagonal structure of $Y$ is taken into account explicitly in the algorithm.

The relation~\eqref{eq:Yyrelation} is an equality between polynomials. For numerical analysis, it needs to be converted into an equality between vectors. This is done by evaluating~\eqref{eq:Yyrelation} at a set of sample points $x_k^{(j)}$, and additionally multiplying these relations by scaling factors $s_k^{(j)}$, where $k=0, \ldots, d_j$. 

The matrices $A_p$ are therefore labelled by $p=(j,r,s,k)$ and are block diagonal similarly to $Y$. Explicitly, we have
\be\label{eq:Aexpression}
	A_{(j,r,s,k)} = \underbrace{0\oplus\cdots\oplus 0}_{2(j-1)\text{ blocks}}
	\oplus
	(s^{(j)}_kQ_{\de_{j1}}^{(j)}(x^{(j)}_k)\otimes E^{rs})
	\oplus 
	(s^{(j)}_kx^{(j)}_kQ_{\de_{j2}}^{(j)}(x^{(j)}_k)\otimes E^{rs})
	\oplus 0\oplus \cdots\oplus 0.
\ee
Explicit expressions for $B$ and $c$ can be found in~\cite{Simmons-Duffin:2015qma}.

\texttt{SDPB} solves the dual SDP~\eqref{eq:dualSDP} simultaneously with a SDP dual to it, referred to as the ``primal SDP''. It is defined as follows,
\be
	&\text{minimize  }c\.x\text{  over  }x\in \R^P,\, X\in \cS^K\nn\\
	&\text{such that  }X=\sum_{p=1}^P A_px_p, \quad B^Tx = b,\quad X\succeq 0.
\ee
Note that $X$ inherits the block-diagonal structure of the matrices $A_p$.

We will not need the full details of the~\texttt{SDPB} algorithm except for the following. The algorithm starts with $X,Y$ proportional to the identity matrix. It then proceeds to improve the values of $X,Y,x,y$ in an iterative fashion. At each iteration, it constructs the ``Schur complement'' matrix $S$ defined as
\be
	S_{ij} = \Tr(A_i X^{-1} A_j Y).
\ee
A Cholesky decomposition $S=LL^T$ of $S$ is computed (it can be verified that $S$ is symmetric and positive-definite) and $L^{-1}$ is used in the computation. This means that the condition number of $S$ has an important effect on the numerical stability of the algorithm. If $S$ is ill-conditioned, the calculation needs to be performed at higher numerical precision.

It is therefore desirable to optimize any user-defined choices to minimize the condition number of $S$. Estimating the condition number of $S$ is not trivial due to its dependence on the undetermined variables $X,Y$. As a proxy, we can try minimizing the condition number of $S$ at the first iteration, when $X$ and $Y$ are proportional to the identity matrix. Since the condition number is independent of normalization, it is enough to consider the case $X=Y=1$, i.e.\
\be
	S_{ij}=\Tr(A_i A_j).
\ee
It is easy to check that~\eqref{eq:Aexpression} implies
\be
	S_{(j,r,s,k),(j',r',s',k')} =& \frac{1}{2}\de_{jj'}\p{\de_{rr'}\de_{ss'}+\de_{rs'}\de_{sr'}}s_k^{(j)}s_{k'}^{(j)}
	\nn\\
	&\x\p{
		\Tr(Q_{\de_{j1}}^{(j)}(x^{(j)}_k)Q_{\de_{j1}}^{(j)}(x^{(j)}_{k'}))+
		x_k^{(j)}x_{k'}^{(j)}\Tr(Q_{\de_{j2}}^{(j)}(x^{(j)}_k)Q_{\de_{j2}}^{(j)}(x^{(j)}_{k'}))
	}.
\ee
Taking into account also the definition~\eqref{eq:Qdefn} we see that it is enough to focus on condition numbers of the matrices $\tl S^{(j)}$ defined as
\be\label{eq:tlS}
	\tl S^{(j)}_{kk'}\equiv  s_k^{(j)}s_{k'}^{(j)}(\vec q^{\,(j)}_{\de_{j1}}(x^{(j)}_k)\.\vec q^{\,(j)}_{\de_{j1}}(x^{(j)}_{k'}))^2
	+s_k^{(j)}s_{k'}^{(j)}x_k^{(j)}x_{k'}^{(j)}(\vec q^{\,(j)}_{\de_{j2}}(x^{(j)}_k)\.\vec q^{\,(j)}_{\de_{j2}}(x^{(j)}_{k'}))^2,
\ee
where $\vec q^{\,(j)}_{\de}(x)=(q_0^{(j)}(x),\ldots,q_\de^{(j)}(x))$.

\subsection{Choice of bilinear bases and sample scalings}
\label{sec:bilinearbasis}

We are now ready to describe the choice of the sample scalings $s_k^{(j)}$ and the bilinear bases $\vec q^{(j)}_\de(x)$ that we have implemented in \texttt{SDPB 3.1}. Although our choice has some motivation that we briefly discuss below, it was ultimately determined by numerical experimentation while trying to improve the initial condition number of the matrix $S$, or equivalently of the matrices~$\tl S^{(j)}$. In this section we ignore some technical implementation details which we defer to section~\ref{sec:implementation}.

The sample points $x_k^{(j)}$ are set to the optimal density interpolation nodes, which are determined as described in section~\ref{sec:optimalnodes}. Following the discussion in section~\ref{sec:errornorm}, we use the weight function $\mu^{(j)}(x)$ given in~\eqref{eq:sdpbprefactor} to define the error norm. This procedure requires some care when poles with small $p_{i,j}$ are present, which we will explain in section~\ref{sec:implementation}. The number of sample points is determined by the maximal degree $d_j$ of the polynomials appearing in the matrices $M_j^k(x)$.

We set the sample scalings $s_k^{(j)}$ to 
\begin{align}
	s_k^{(j)} = \mu^{(j)}(x_k^{(j)}).
\end{align}
For the bilinear bases, we make different choices for the two terms in~\eqref{eq:Yyrelation}. For the first term, i.e.\ for the matrices $Q^{(j)}_{\de_{j1}}$ we choose $q_0^{(j)}(x),\ldots,q_{\de_{j1}}^{(j)}(x)$ to be the polynomials orthonormal with respect to the inner product
\begin{align}\label{eq:inner1}
	\<p|q\>_1 = \sum_{k=0}^{d_j} \mu^{(j)}(x_k) p(x_k) q(x_k).
\end{align}
For the second term, i.e.\ for the matrices $Q^{(j)}_{\de_{j2}}$ we choose $q_0^{(j)}(x),\ldots,q_{\de_{j2}}^{(j)}(x)$ to be the polynomials orthonormal with respect to the inner product
\begin{align}\label{eq:inner2}
	\<p|q\>_2 = \sum_{k=0}^{d_j} x_k\mu^{(j)}(x_k) p(x_k) q(x_k).
\end{align}
It was not possible to specify two different choices of bilinear bases for the two terms~\eqref{eq:Yyrelation} in~\texttt{SDPB 3.0}. We have added this option in~\texttt{SDPB 3.1}.

Let us now discuss some intuition behind these choices. The sample points $x_k^{(j)}$ are set to the optimal density interpolation nodes so that they have a small Lebesgue constant for the error norm defined by $\mu^{(j)}(x)$. This minimizes some condition numbers related to the implementation of the equation~\eqref{eq:Yyrelation}. Indeed, we encode this equation by evaluating it at the sample points $x_k^{(j)}$ and  multiplying by the sample scalings $s_k^{(j)}= \mu^{(j)}(x_k^{(j)})$. In other words, we encode it as an equality in $\R^{d_j+1}$ by applying the isomorphism $\f$ of section~\ref{sec:interpolation}. There, we saw that the condition number of $\f$ is bounded by the Lebesgue constant. We therefore expect that using the optimal density interpolation nodes as sample points and the above sample scalings gives the most numerically stable encoding of~\eqref{eq:Yyrelation}.

The choice of bilinear bases is motivated by the prior expectation that polynomials orthonormal with respect to the inner product
\begin{align}
	\<p|q\>_\text{cont} = \int_0^\oo dx \mu^{(j)}(x)p(x)q(x)
\end{align}
should give a reasonable basis. In particular, these were the polynomials that were used by \texttt{SDPB 3.0}. Empirically, we found that the similar inner products~\eqref{eq:inner1} and~\eqref{eq:inner2} give better initial condition numbers for $S$,\footnote{Note that while~\eqref{eq:inner1} and~\eqref{eq:inner2} look somewhat like Riemann sums for $\<p|q\>_\text{cont}$, they are missing the factor of the form $x_{k+1}-x_k$ that a Riemann sum would have.} and are easier to compute in practice.

\subsection{Implementation details}
\label{sec:implementation}

In this section we discuss some technical details of how the choices discussed above are implemented in~\texttt{SDPB 3.1}.

The discussion below concerns the scenario where the sample scalings, sample points, and bilinear bases are computed by~\texttt{pmp2sdp} (the user has the ability to override these values). In this scenario, the user provides the following data for each positivity constraint, i.e.\ for each value of $j$:
\begin{itemize}
	\item the polynomial matrices $\tl M^k_j(x)$,
	\item the prefactor 
	\begin{align}
	\label{eq:theformofprefactor}
		\tl\mu^{(j)}(x) &= \frac{\tl c_j\tl r_j^x}{\prod_{i=0}^{\tl K_j} (x-\tl p_{j,i})^{\tl\a_{j,i}}},
	\end{align}
	\item the ``reduced" prefactor
	\begin{align}
	\label{eq:details_mu}
		\mu^{(j)}(x) &= \frac{c_jr_j^x}{\prod_{i=0}^{K_j}(x-p_{j,i})^{\a_{j,i}}}.
	\end{align}
\end{itemize}
We assume that $p_{j,i},\tl p_{j,i}\leq 0$. The effect of the reduced prefactor is that \texttt{pmp2sdp} will construct a set of interpolating polynomial matrices $M^k_j(x)$ such that
\begin{align}\label{eq:approximation}
	\tl \mu^{(j)}(x) \tl M^k_j(x)\approx 
	\mu^{(j)}(x) M^k_j(x),
\end{align}
minimizing the error with respect to the uniform $\sup$-norm in this equation. Equivalently, the reduced prefactor defines the weight function $\mu^{(j)}(x)$,
and the $M^k_j(x)$ interpolate the functions
\begin{align}
\label{eq:thingweinterpolate}
	\frac{\tl \mu^{(j)}(x)}{\mu^{(j)}(x)}\tl M^k_j(x).
\end{align}
The sample points are set to the optimal density interpolation nodes and minimize the Lebesgue constant with respect to the error norm defined by $\mu^{(j)}(x)$ on $[0,+\oo)$.

The matrix functions
\begin{align}
	\tl\mu^{(j)}(x) \tl M^k_j(x)
\end{align}
should be seen as high-precision approximations to the true matrix functions. For instance, they might be constructed using the block approximations output by \texttt{scalar\_blocks} with parameters \texttt{order} and \texttt{poles} (the latter is also known as kept pole order) equal and large. The polynomial degree $d_j$ used for interpolating polynomials $M^k_j(x)$ is such that both sides of~\eqref{eq:approximation} have the same growth at infinity, i.e.
\begin{align}
	d_j = \sum_{i=0}^{K_j}\a_{j,i}-\sum_{i=0}^{\tl K_j}\tl \a_{j,i} + \max_k \deg \tl M^k_j(x).
\end{align}
Therefore, the user can control the approximation degree through the total number of poles provided in the reduced prefactor.

Some care is required when poles at small $|x|$ are present. Note first that the mathematical recipe in section~\ref{sec:optimalnodes} for computing interpolation nodes continues to work, even when there are poles close to or at $x=0$. For example, if there is a pole of degree $k$ at zero, then the Bohr-Sommerfeld quantization condition will yield $k$ degenerate interpolation nodes at zero, followed by some collection of nonzero interpolation nodes. However, instead of allowing degenerate interpolation nodes, we found it simpler to choose non-optimal but non-degenerate nodes near a pole, as follows. We define a parameter
\be
\texttt{smallPoleTheshold=}&10^{-10}.
\ee
We count poles $p$ in $x$ (with multiplicity) whose locations satisfy $|p|<\texttt{smallPoleTheshold}$. We call these ``small poles." Suppose there are $k$ small poles. We then compute Bohr-Sommerfeld interpolation nodes $\{x_1,\dots,x_N\}$ as usual, and replace the first $k$ (possibly degenerate) nodes with uniformly spaced nodes between 0 and the $k+1$-st node. That is, we choose the sample points
\be
\left\{\frac{x_{k+1}}{k},\frac{2x_{k+1}}{k},\dots,\frac{(k-1)x_{k+1}}{k},x_{k+1},\dots,x_N\right\}.
\ee

In addition to computing sample points, we must also compute ``sample scalings" and ``reduced sample scalings," which are the values of the prefactor $\tl\mu^{(j)}(x)$ and reduced prefactor $\mu^j(x)$ at the sample points. If any of the sample points are close to a pole, these values must be regulated. To do so, we define a parameter
\begin{align}
	\texttt{minPoleDistance=}&10^{-16},
\end{align}
and in evaluating $\tl\mu^{(j)}(x)$ and $\mu^j(x)$, we replace each factor in the denominator as follows:
\be
(x-p) &\longrightarrow \max(x-p, \texttt{minPoleDistance}).
\ee
Note that the actual values of the sample scalings only matter for numerical stability, so this modification does not affect correctness of the resulting SDP. One important point is that the $M_j^k(x)$ interpolate the functions (\ref{eq:thingweinterpolate}), which involve a ratio of the prefactor over the reduced prefactor $\tl\mu^{(j)}(x)/\mu^{(j)}(x)$. Thus, it is crucial that we regulate the prefactor and reduced prefactor in the same way, so that the effects of the regulator cancel in this ratio. If a regulated pole were present in $\tl\mu^{(j)}(x)$ but not $\mu^{(j)}(x)$ (or vice-versa), then it would lead to errors. Note that our choice of keeping the rightmost $n_\mathrm{poles}$ in the reduced prefactor avoids this problem, as long as $n_\mathrm{poles}$ is larger than the total degree of poles within $\texttt{minPoleDistance}$ of $x=0$.

\section{Numerical tests}
\label{sec:numericaltests}

To test our new conformal block approximations and the solver choices described in section~\ref{sec:conditionnumbers}, we studied the following bootstrap problem: compute a lower bound on the central charge $c_T$ in a 3d CFT containing a scalar $\f$ with dimension $\De_\f=0.5181489$, using crossing symmetry of the four-point function $\<\f\f\f\f\>$ \cite{Rattazzi:2010gj,Poland:2010wg}. We computed this bound using two setups:
\begin{enumerate}
\item {\bf (Pole shifting)} This setup uses the ``pole-shifting" method described in \cite{Kos:2013tga} and the default bilinear bases, sample points, and sample scalings in {\tt SDPB} 3.0. The conformal block approximations in this setup depend on the parameter $\texttt{keptPoleOrder}$.\footnote{The code for setup 1 is at:  \url{https://gitlab.com/davidsd/scalars-3d/-/blob/47440dc1b60f9cbe92519a13d933d05c79d1a64c/src/Projects/Scalars3d/SingletScalarBlocks3d2020.hs\#L184}.}
\item {\bf (Interpolation)} This setup uses the optimal density interpolation nodes as sample points and the choice of bilinear bases and sample scalings described in this work (and implemented in {\tt SDPB} 3.1). The conformal block approximations in this setup depend on the parameter $n_\mathrm{poles}$: the number of poles in the reduced prefactor $\mu^{(j)}(x)$.\footnote{The code for setup 2 is at: \url{https://gitlab.com/davidsd/scalars-3d/-/blob/47440dc1b60f9cbe92519a13d933d05c79d1a64c/src/Projects/Scalars3d/SingletScalarBlocks3d2020.hs\#L272}, using {\tt SDPB} revision \url{https://github.com/davidsd/sdpb/tree/446b2b9d07df8984ecfd6a66edb491261a9b1f65}.}

In order to use this setup with {\tt SDPB} 3.1, we first generate positivity constraints in the usual form
\be
\label{eq:ourpositivitycondition}
\tl \mu^{(j)}(x) \sum_k z_k \tl M^k_j(x) &\succeq 0,
\ee
where $\tl \mu^{(j)}(x)\tl M^k_j(x)$ are approximations to derivatives of conformal blocks obtained by truncating the $r$-expansion at some {\tt order} --- with no pole-shifting. The $z_k$ are coefficients of the functional that the solver will determine. The ``prefactor" $\tl \mu^{(j)}(x)$ has the damped rational form (\ref{eq:theformofprefactor}) with $r_j=r_*=3-2\sqrt 2$, where the number of poles appearing might be quite large (it is roughly between $\tfrac{3}{2}\mathtt{order}$ and $\tfrac{5}{2}\mathtt{order}$, depending on the spin of the block). Finally, $\tl M^k_j(x)$ are matrices of polynomials.

We now define the reduced prefactor $\mu^{(j)}(x)$ to have the same form as $\tl \mu^{(j)}(x)$, but only including the $n_\mathrm{poles}$ rightmost poles, taking into account multiplicities. For example, with $n_\mathrm{poles}=4$, we might have
\be
\tl\mu^{(j)}(x) = \frac{r_*^x}{(x+1)^2(x+2)(x+3)^3(x+4)},\qquad \mu^{(j)}(x)=\frac{r_*^x}{(x+1)^2(x+2)(x+3)}.
\ee
Here, the list of poles of $\tl\mu^{(j)}(x)$ is (including multiplicities): $\{-1,-1,-2,-3,-3,-3,-4\}$. Truncating this to length $n_\mathrm{poles}=4$ gives $\{-1,-1,-2,-3\}$, which are the poles of $\mu^{(j)}(x)$.

We encode the positivity condition (\ref{eq:ourpositivitycondition}) using a ``positive matrix with prefactor" as described in the {\tt SDPB} 3.1 manual, where $\mathtt{prefactor}=\tl\mu^{(j)}(x)$ and $\mathtt{reducedPrefactor}=\mu^{(j)}(x)$. We omit custom choices of {\tt samplePoints}, {\tt sampleScalings}, {\tt reducedSampleScalings}, {\tt bilinearBasis\_0}, and {\tt bilinearBasis\_1}, so these will default to the settings described in this work.
\end{enumerate}
In both setups, we compute conformal blocks using {\tt blocks\_3d} up to ${\tt order}=80$ in the $r$-expansion. In setup 2, we turn off pole-shifting by setting $\mathtt{keptPoleOrder}=\mathtt{order}$. Our parameters for the two setups are summarized in table~\ref{tab:setups}.

\begin{table}
\centering
\begin{tabular}{c|c|c}
parameter & setup 1 & setup 2 \\
\hline
{\tt nmax} & 10 & 10 \\
\hline
{\tt order} & 80 & 80 \\
\hline
{\tt keptPoleOrder} & 20,30,40,50,60,70,80 & 80 \\
\hline
$n_\mathrm{poles}$ & -- & 10,20,30,40,50,60,70,80,90,100 \\
\hline
{\tt precision} & 1536 & 1280 \\
\hline
{\tt dualityGapThreshold} & $10^{-100}$ & $10^{-100}$ \\
\hline
{\tt SDPB} version & 3.0.0 & 3.1
\end{tabular}
\caption{\label{tab:setups}Parameters for computing central charge bounds using setups 1 and 2. Note that when we attempted to use setup 1 with $\texttt{keptPoleOrder}\geq 70$ and $\mathtt{precision}=1280$, the solver failed with {\tt MaxComplementarityExceeded}. Thus, we increased precision to 1536 for the sake of this test. The parameter ${\tt nmax}$ is related to the derivative order $\Lambda$ by $\Lambda=2\,\texttt{nmax}-1$. Note that $\mathtt{precision}=1280$ in setup 2 is much higher than necessary to achieve accurate bounds. We use large precision simply because we would like to isolate the effects of changing $n_\mathrm{poles}$.
}\end{table}

We found that the primal objectives for setup 1 with $\texttt{keptPoleOrder}\geq 80$ and setup 2 with $n_\mathrm{poles}\geq 90$ all coincided, within the value of the duality gap threshold $10^{-100}$. We define the ``most accurate bound" as the value of the primal objective in these cases. We define the accuracy of other bounds by taking the difference between their primal objective and the most accurate bound.

\begin{figure}
    \centering
    \includegraphics[width=0.9\textwidth]{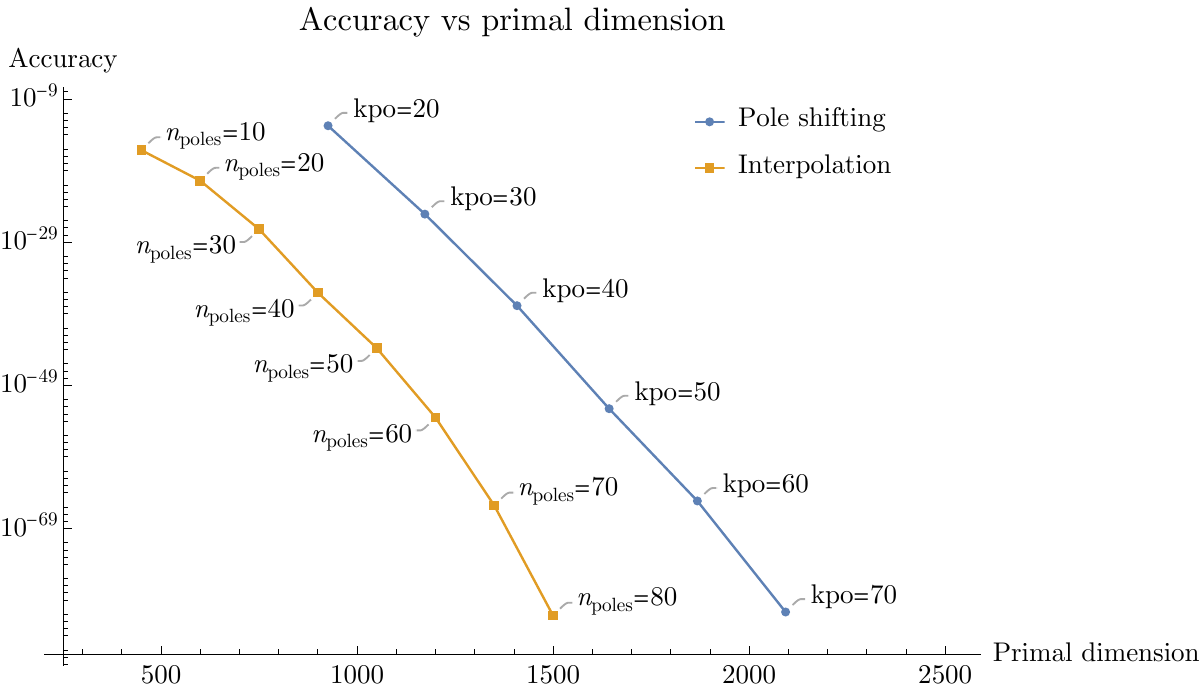}
    \caption{A comparison of \texttt{SDPB} solution accuracy vs.\ primal dimension for the two setups. Here and in the plots below, we use the shorthand notation $\mathtt{keptPoleOrder}\longrightarrow\mathrm{kpo}$.}
    \label{fig:accuracy_vs_dim}
\end{figure}

We can quantify the size of the matrices inside {\tt SDPB} using the ``primal dimension," i.e.\ the dimension of the primal objective vector $c$. A larger primal dimension corresponds to larger-degree polynomials, larger matrices, and generally worse memory usage and performance. In figure~\ref{fig:accuracy_vs_dim}, we plot accuracy vs.\ primal dimension for the two setups. We see that setup 2 achieves much better accuracy at the same value of the primal dimension. For example, setup 2 is more accurate by a factor of $\approx 10^{-30}$ at primal dimension 1300. Furthermore, accuracy improves with increasing primal dimension faster in setup 2 than in setup 1.\footnote{\label{note:polechoice}The bounds in setup 2 could actually be made {\it even more\/} accurate by a simple change to {\tt blocks\_3d}, but this change would only affect single-correlator bounds. Currently, {\tt blocks\_3d} creates damped rational approximations that include the full set of poles that appear in Zamolodchikov-type recursion relations for generic external dimensions $\De_i$, without checking whether some of the residues vanish. For conformal blocks with equal external dimensions, some of the residues {\it do\/} vanish, and consequently the damped rationals computed by {\tt blocks\_3d} have extra poles. In accordance with the discussion of the choice of poles at the end of section~\ref{sec:errornorm}, including these extra poles among the $n_\mathrm{poles}$ poles of the reduced prefactor is inefficient and reduces the accuracy of approximation. Thus, figure~\ref{fig:accuracy_vs_dim} is more representative of a general mixed correlator system where generically all conformal block poles have nonzero residues. Additionally, because pole-shifting in {\tt blocks\_3d} also keeps all the poles for equal external dimensions, our comparison between pole-shifting and interpolation is still reasonable.}

\begin{figure}
    \centering
    \includegraphics[width=0.9\textwidth]{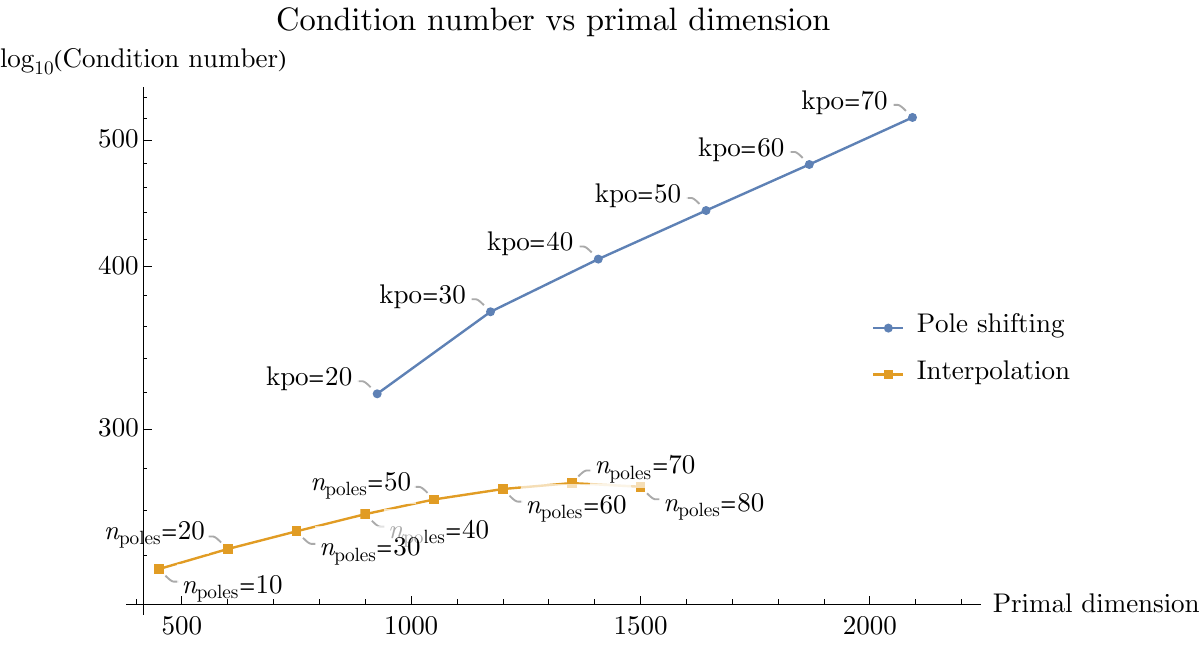}
    \caption{Maximum condition number during an \texttt{SDPB} run for the two setups. The maximum condition number is typically achieved by a block of the matrix $S$.}
    \label{fig:cond_number_vs_dim}
\end{figure}

\begin{figure}
    \centering
    \includegraphics[width=0.9\textwidth]{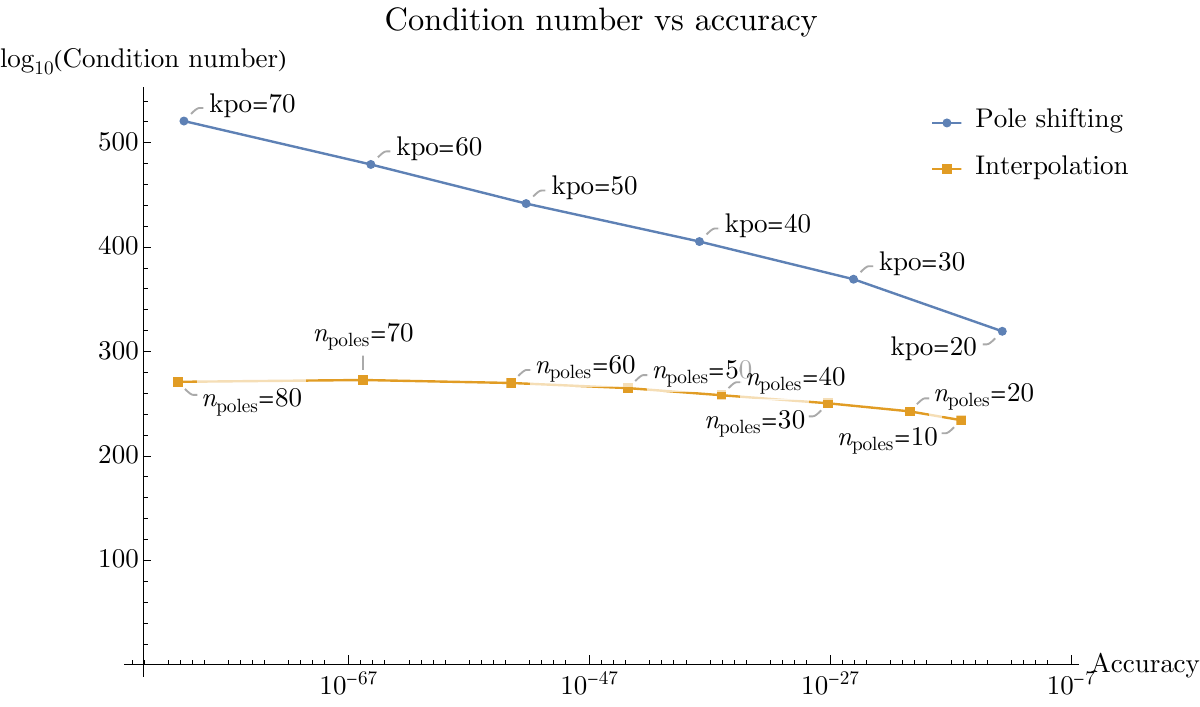}
    \caption{Maximum condition number vs.\ accuracy for the two setups.  This plot is simply a different way of viewing data from figures~\ref{fig:accuracy_vs_dim} and \ref{fig:cond_number_vs_dim}. We see that we can systematically improve accuracy of bootstrap bounds by increasing $n_\mathrm{poles}$, without significantly increasing condition numbers inside {\tt SDPB}. By contrast, increasing {\tt keptPoleOrder} leads to large increases in condition numbers that must be dealt with by increasing {\tt precision}.}
    \label{fig:cond_number_vs_accuracy}
\end{figure}

In figure~\ref{fig:cond_number_vs_dim}, we show how condition numbers inside {\tt SDPB} behave in the two setups. We define the max condition number for an {\tt SDPB} run as the largest condition number of any internal matrix that must be inverted and/or Cholesky-decomposed during a solver run (namely, the matrices $X,Y,S$, and $Q$). In practice, the largest condition number typically comes from one of the blocks of the Schur complement matrix $S$. Our new choice of bilinear basis described in section~\ref{sec:bilinearbasis} was designed to minimize the condition number of $S$ at the beginning of an {\tt SDPB} run. However, with our new choices, we also observe significantly reduced condition numbers throughout a run. Another important benefit of setup 2 is that condition numbers no longer grow significantly with primal dimension. Thus, in setup 2 we are free to improve the accuracy of conformal block approximations without having to run the solver at higher precision. (The minimal precision required to avoid crashing {\tt SDPB} is approximately $2\log_{10}(\textrm{condition number})$). This addresses one of the main difficulties we encountered when using pole-shifting in the Ising+stress tensors bootstrap \cite{Chang:2024whx}. We give another way to visualize this result in figure~\ref{fig:cond_number_vs_accuracy} where we plot maximum condition number vs.\ accuracy.

\begin{figure}
    \centering
    \includegraphics[width=\textwidth]{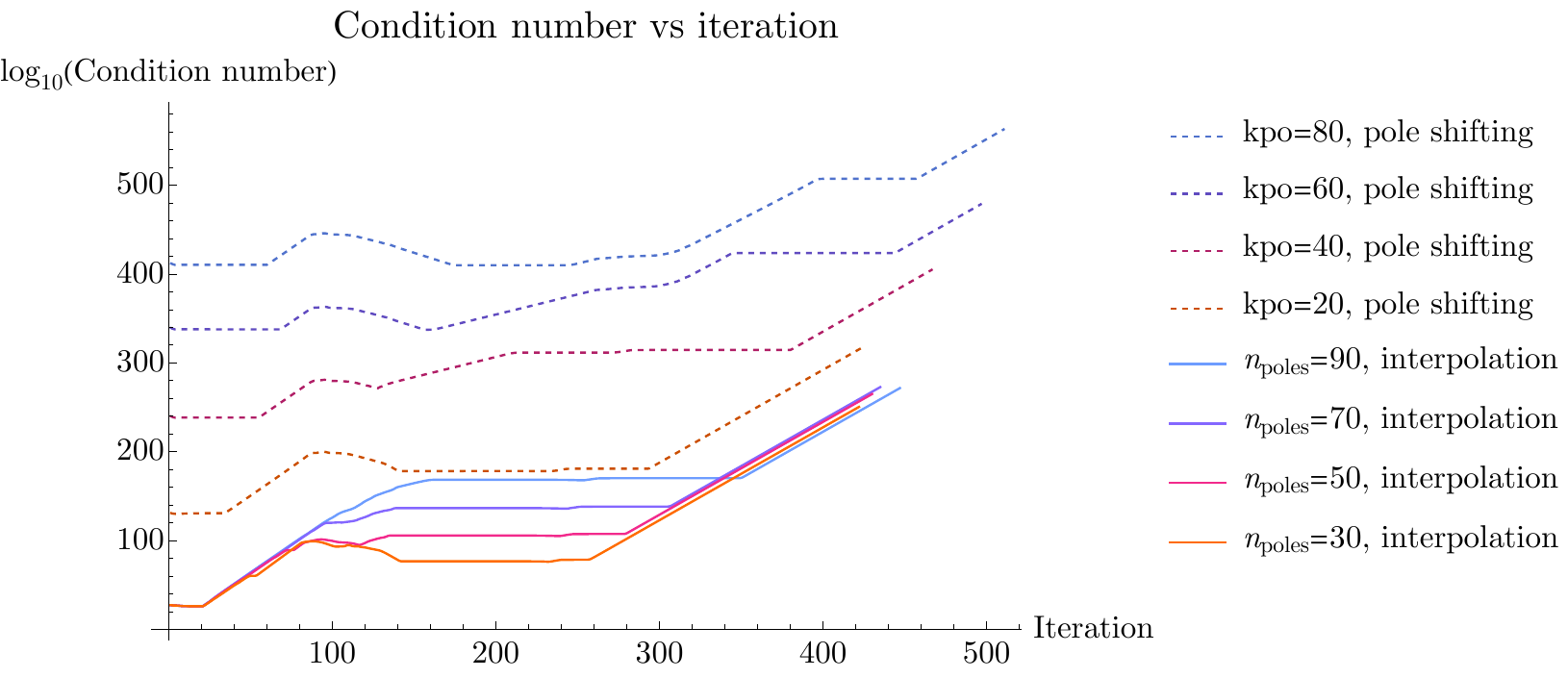}
    \caption{Maximum condition number vs.\ \texttt{SDPB} solver iteration for different approximation schemes. }
    \label{fig:cond_number_vs_iteration}
\end{figure}

It is also instructive to look at how condition numbers evolve during an {\tt SDPB} run. In figure~\ref{fig:cond_number_vs_iteration}, we plot max condition number vs.\ iteration. The condition numbers in setup 1 are large at the very beginning (with worse condition numbers at higher {\tt keptPoleOrder}), and increase over the course of a run. By contrast, the condition numbers in setup 2 start small, and are consistent for different values of $n_\mathrm{poles}$. They then begin to increase, plateauing in the middle of the run at a location that depends on $n_\mathrm{poles}$, and finally growing again at the end of the run, but still remaining far below the condition numbers in setup 1.\footnote{Note that large condition numbers at the end of a solver run are essentially guaranteed in a primal-dual interior point solver like {\tt SDPB}. The reason is that the solver moves towards a solution of the equation $X Y=\mu I$, where $X,Y$ are positive-semidefinite, and $\mu$ is a parameter that is going to zero. In typical SDP's, both $X$ and $Y$ will have some very small eigenvalues near optimality. It would be interesting to understand theoretically how large we should expect final condition numbers to be.}

\section{Future directions}
\label{sec:future}

We have argued that to obtain accurate numerical conformal bootstrap bounds, we should approximate the optimal functional $\alpha_\mathrm{opt}(F_{\De,j})$ as closely as possible in the regime where it has interesting structure. For functionals built from a finite number of derivatives of the crossing equation, the interesting region is a finite range of $\De$ where the functional exhibits ``wiggles," beyond which it decays exponentially. This observation leads to a choice of norm $\|\cdot\|_{\mathrm{bootstrap},j}$ that measures the size of contributions in this region, and a corresponding optimal density of interpolation nodes. In practice, polynomial interpolation using optimal density nodes leads to accurate, exponentially convergent approximations for conformal blocks and optimal functionals. More importantly, it leads to accurate numerical bootstrap bounds with fewer computational resources, both by increasing accuracy of approximations and improving condition numbers inside {\tt SDPB}.

For simplicity, we constructed $\|\cdot\|_{\mathrm{bootstrap},j}$ by multiplying conformal blocks by the exponential factor $4^{-\De}$, which can be motivated either by considering the asymptotics of spin-2 dispersive functionals (\ref{eq:dispersivefunctionalasymptotics}), or by considering the asymptotic behavior of MFT OPE coefficients (\ref{eq:mftasympt}). However, we did not take into account power-law corrections $\De^\#$ to these asymptotics. This was largely for practical reasons: the exponential factor $4^{-\De}$ is the same for arbitrary external operators, and so is easier to work with in mixed correlator systems. Furthermore, a power law factor like $\De^{\frac 1 2 + 4\De_\f}$ does not change the optimal density interpolation nodes or condition numbers in the solver by a large amount. A more subtle point is that in mixed correlator problems, multiplying different matrix elements of the functional by different powers of $\De$ would create a mismatch between the large-$\De$ asymptotics of individual blocks and the large-$\De$ asymptotics of our ansatz for them. Whether this mismatch affects the accuracy of the resulting bootstrap bounds deserves further study.

Optimal functionals built from linear combinations of $N$ derivatives of conformal blocks generically exhibit at most $N/2$ double zeros, by dimension counting. In practice, most of these zeros, and thus most of the structure in the optimal functional, is concentrated at small spin, see e.g.\ \cite{Simmons-Duffin:2016wlq}. Thus the action of an extremal functional $\alpha_\mathrm{opt}(F_{\De,j})$ typically exhibits less and less structure at larger and larger $j$. This suggests that it might be fruitful to reduce the number of interpolation nodes used for approximating blocks at large $j$. This would further reduce the degree of polynomials in {\tt SDPB}, potentially significantly improving performance. It would be interesting to explore this possibility in the future.

In this work, we chose a new convention for which poles to include in the ansatz (\ref{eq:thingwewanttodoagain}) for conformal block approximations. Previous works using pole-shifting \cite{Kos:2013tga} had included poles with the largest residues --- specifically, they restricted to poles $\De_A$ whose corresponding order $n_A$ in Zamolodchikov-type recursion relations \cite{Kos:2013tga,Kos:2014bka,Penedones:2015aga} was less than a parameter {\tt keptPoleOrder}. In this work, we instead simply include the  $n_\mathrm{poles}$ rightmost poles of the conformal block. We have observed that not including the rightmost poles in the block makes the interpolation worse, since it decreases the analyticity domain of the function we want to interpolate. We have also found that for identical external scalars, including only the poles with nonzero residues could improve the interpolation. It would be interesting to explore the choice of poles more systematically in the future. 
We also have not fully explored the possible space of bilinear bases. Perhaps this choice can be optimized as well to improve condition numbers not only at the beginning of an {\tt SDPB} run, but throughout the run.

The semidefinite programming approach to the bootstrap consists of approximating the action of functionals in terms of manifestly positive sums of products of bilinear bases. When the functionals are approximated in terms of polynomials, it is natural for the bilinear bases to be polynomials as well. However, this choice is not necessary. We could consider approximating functionals in terms of a different kind of bilinear basis. For example, it may be interesting to consider ``analytic" functionals, such as those studied in \cite{Mazac:2016qev,Mazac:2018mdx,Mazac:2018ycv,Paulos:2019gtx,Penedones:2019tng,Caron-Huot:2020adz,Caron-Huot:2022sdy,Ghosh:2023onl}. If one could construct a reasonable bilinear basis for analytic functionals, and a corresponding set of sample points, then one could potentially apply the usual semidefinite programming machinery, using equality between functionals and products of bilinear basis elements at sample points as the primal constraints.

\section*{Acknowledgements}

We thank Rajeev Erramilli, Alexandre Homrich, Aike Liu, and Matthew Mitchell for collaboration on the Ising stress tensors project that motivated this work, and for helpful discussions.
CHC is supported by a Kadanoff fellowship at the University of Chicago.
DP is supported by Simons Foundation grant 488651 (Simons Collaboration on the Nonperturbative Bootstrap) and DOE grant DE-SC0017660.
The work of PK was funded by UK Research and Innovation (UKRI) under the UK government's Horizon Europe funding Guarantee [grant number EP/X042618/1] and the Science
and Technology Facilities Council [grant number ST/X000753/1].
DSD and VD are supported in part by Simons Foundation grant 488657 (Simons Collaboration on the Nonperturbative Bootstrap) and the U.S. Department of Energy, Office of Science, Office of High Energy Physics, under Award Number DE-SC0011632. 
This work used the Expanse cluster at the San Diego Supercomputing Center (SDSC) through allocation
PHY190023 from the Advanced Cyberinfrastructure Coordination Ecosystem: Services
\& Support (ACCESS) program, which is supported by National Science Foundation
grants \#2138259, \#2138286, \#2138307, \#2137603, and \#2138296.

\appendix

\bibliographystyle{JHEP}
\bibliography{refs}

@inproceedings{Marino:2004eq,
    author = "Marino, Marcos",
    title = "{Les Houches lectures on matrix models and topological strings}",
    eprint = "hep-th/0410165",
    archivePrefix = "arXiv",
    reportNumber = "CERN-PH-TH-2004-199",
    month = "10",
    year = "2004"
}

@misc{muskhelishvili1953singular,
  title={Singular Integral Equations},
  author={Muskhelishvili, NI},
  year={1953},
  publisher={Reprinted by Dover Publications: New York}
}

@book{saff2013logarithmic,
  title={Logarithmic potentials with external fields},
  author={Saff, Edward B and Totik, Vilmos},
  volume={316},
  year={2013},
  publisher={Springer Science \& Business Media}
}

@incollection {BrutmanLebesgue,
	AUTHOR = {Brutman, L.},
	TITLE = {Lebesgue functions for polynomial interpolation---a survey},
	NOTE = {The heritage of P. L. Chebyshev: a Festschrift in honor of the
	70th birthday of T. J. Rivlin},
	JOURNAL = {Ann. Numer. Math.},
	FJOURNAL = {Annals of Numerical Mathematics},
	VOLUME = {4},
	YEAR = {1997},
	NUMBER = {1-4},
	PAGES = {111--127},
	ISSN = {1021-2655},
	MRCLASS = {41A05 (41-02)},
	MRNUMBER = {1422674},
	MRREVIEWER = {Eduard\ Belinskii},
}

@article{Simmons-Duffin:2016wlq,
  author        = {Simmons-Duffin, David},
  title         = {{The Lightcone Bootstrap and the Spectrum of the 3d Ising CFT}},
  eprint        = {1612.08471},
  archiveprefix = {arXiv},
  primaryclass  = {hep-th},
  doi           = {10.1007/JHEP03(2017)086},
  journal       = {JHEP},
  volume        = {03},
  pages         = {086},
  year          = {2017}
}

@article{Poland:2018epd,
  author        = {Poland, David and Rychkov, Slava and Vichi, Alessandro},
  title         = {{The Conformal Bootstrap: Theory, Numerical Techniques, and Applications}},
  eprint        = {1805.04405},
  archiveprefix = {arXiv},
  primaryclass  = {hep-th},
  doi           = {10.1103/RevModPhys.91.015002},
  journal       = {Rev. Mod. Phys.},
  volume        = {91},
  pages         = {015002},
  year          = {2019}
}

@unpublished{interpolation,
  title = {{Improved Polynomial Interpolation for Conformal Blocks, to appear}}
}

@article{Fitzpatrick:2012yx,
  author        = {Fitzpatrick, A. Liam and Kaplan, Jared and Poland, David and Simmons-Duffin, David},
  title         = {{The Analytic Bootstrap and AdS Superhorizon Locality}},
  eprint        = {1212.3616},
  archiveprefix = {arXiv},
  primaryclass  = {hep-th},
  doi           = {10.1007/JHEP12(2013)004},
  journal       = {JHEP},
  volume        = {12},
  pages         = {004},
  year          = {2013}
}

@article{Komargodski:2012ek,
  author        = {Komargodski, Zohar and Zhiboedov, Alexander},
  title         = {{Convexity and Liberation at Large Spin}},
  eprint        = {1212.4103},
  archiveprefix = {arXiv},
  primaryclass  = {hep-th},
  doi           = {10.1007/JHEP11(2013)140},
  journal       = {JHEP},
  volume        = {11},
  pages         = {140},
  year          = {2013}
}

@article{Rychkov:2015lca,
    author = "Rychkov, Slava and Yvernay, Pierre",
    title = "{Remarks on the Convergence Properties of the Conformal Block Expansion}",
    eprint = "1510.08486",
    archivePrefix = "arXiv",
    primaryClass = "hep-th",
    reportNumber = "CERN-PH-TH-2015-253",
    doi = "10.1016/j.physletb.2016.01.004",
    journal = "Phys. Lett. B",
    volume = "753",
    pages = "682--686",
    year = "2016"
}

@article{Caron-Huot:2021enk,
    author = "Caron-Huot, Simon and Mazac, Dalimil and Rastelli, Leonardo and Simmons-Duffin, David",
    title = "{AdS bulk locality from sharp CFT bounds}",
    eprint = "2106.10274",
    archivePrefix = "arXiv",
    primaryClass = "hep-th",
    doi = "10.1007/JHEP11(2021)164",
    journal = "JHEP",
    volume = "11",
    pages = "164",
    year = "2021"
}

@article{Chang:2023szz,
    author = "Chang, Cyuan-Han and Landau, Yakov and Simmons-Duffin, David",
    title = "{Spinning dispersive CFT sum rules and bulk scattering}",
    eprint = "2311.04271",
    archivePrefix = "arXiv",
    primaryClass = "hep-th",
    reportNumber = "CALT-TH 2023-043",
    doi = "10.1007/JHEP04(2025)016",
    journal = "JHEP",
    volume = "04",
    pages = "016",
    year = "2025"
}

@article{Qiao:2017lkv,
    author = "Qiao, Jiaxin and Rychkov, Slava",
    title = "{Cut-touching linear functionals in the conformal bootstrap}",
    eprint = "1705.01357",
    archivePrefix = "arXiv",
    primaryClass = "hep-th",
    reportNumber = "CERN-PH-TH-2017-098, CERN-TH-2017-098",
    doi = "10.1007/JHEP06(2017)076",
    journal = "JHEP",
    volume = "06",
    pages = "076",
    year = "2017"
}

@article{Hogervorst:2013sma,
    author = "Hogervorst, Matthijs and Rychkov, Slava",
    title = "{Radial Coordinates for Conformal Blocks}",
    eprint = "1303.1111",
    archivePrefix = "arXiv",
    primaryClass = "hep-th",
    reportNumber = "CERN-PH-TH-2013-043, LPTENS-13-05",
    doi = "10.1103/PhysRevD.87.106004",
    journal = "Phys. Rev. D",
    volume = "87",
    pages = "106004",
    year = "2013"
}

@article{Pappadopulo:2012jk,
    author = "Pappadopulo, Duccio and Rychkov, Slava and Espin, Johnny and Rattazzi, Riccardo",
    title = "{OPE Convergence in Conformal Field Theory}",
    eprint = "1208.6449",
    archivePrefix = "arXiv",
    primaryClass = "hep-th",
    reportNumber = "LPTENS-12-31",
    doi = "10.1103/PhysRevD.86.105043",
    journal = "Phys. Rev. D",
    volume = "86",
    pages = "105043",
    year = "2012"
}

@article{Chang:2024whx,
    author = "Chang, Cyuan-Han and Dommes, Vasiliy and Erramilli, Rajeev S. and Homrich, Alexandre and Kravchuk, Petr and Liu, Aike and Mitchell, Matthew S. and Poland, David and Simmons-Duffin, David",
    title = "{Bootstrapping the 3d Ising stress tensor}",
    eprint = "2411.15300",
    archivePrefix = "arXiv",
    primaryClass = "hep-th",
    reportNumber = "CALT-TH 2024-047",
    doi = "10.1007/JHEP03(2025)136",
    journal = "JHEP",
    volume = "03",
    pages = "136",
    year = "2025"
}

@article{Penedones:2015aga,
    author = "Penedones, Jo{\~a}o and Trevisani, Emilio and Yamazaki, Masahito",
    title = "{Recursion Relations for Conformal Blocks}",
    eprint = "1509.00428",
    archivePrefix = "arXiv",
    primaryClass = "hep-th",
    reportNumber = "IPMU15-0139",
    doi = "10.1007/JHEP09(2016)070",
    journal = "JHEP",
    volume = "09",
    pages = "070",
    year = "2016"
}

@article{Erramilli:2019njx,
    author = "Erramilli, Rajeev S. and Iliesiu, Luca V. and Kravchuk, Petr",
    title = "{Recursion relation for general 3d blocks}",
    eprint = "1907.11247",
    archivePrefix = "arXiv",
    primaryClass = "hep-th",
    reportNumber = "PUPT-2593",
    doi = "10.1007/JHEP12(2019)116",
    journal = "JHEP",
    volume = "12",
    pages = "116",
    year = "2019"
}

@article{Erramilli:2020rlr,
    author = "Erramilli, Rajeev S. and Iliesiu, Luca V. and Kravchuk, Petr and Landry, Walter and Poland, David and Simmons-Duffin, David",
    title = "{blocks{\_}3d: software for general 3d conformal blocks}",
    eprint = "2011.01959",
    archivePrefix = "arXiv",
    primaryClass = "hep-th",
    reportNumber = "CALT-TH 2020-048",
    doi = "10.1007/JHEP11(2021)006",
    journal = "JHEP",
    volume = "11",
    pages = "006",
    year = "2021"
}

@article{Kos:2013tga,
    author = "Kos, Filip and Poland, David and Simmons-Duffin, David",
    title = "{Bootstrapping the $O(N)$ vector models}",
    eprint = "1307.6856",
    archivePrefix = "arXiv",
    primaryClass = "hep-th",
    doi = "10.1007/JHEP06(2014)091",
    journal = "JHEP",
    volume = "06",
    pages = "091",
    year = "2014"
}

@article{Kos:2015mba,
    author = "Kos, Filip and Poland, David and Simmons-Duffin, David and Vichi, Alessandro",
    title = "{Bootstrapping the O(N) Archipelago}",
    eprint = "1504.07997",
    archivePrefix = "arXiv",
    primaryClass = "hep-th",
    reportNumber = "CERN-PH-TH-2015-097",
    doi = "10.1007/JHEP11(2015)106",
    journal = "JHEP",
    volume = "11",
    pages = "106",
    year = "2015"
}

@article{Kos:2016ysd,
    author = "Kos, Filip and Poland, David and Simmons-Duffin, David and Vichi, Alessandro",
    title = "{Precision Islands in the Ising and $O(N)$ Models}",
    eprint = "1603.04436",
    archivePrefix = "arXiv",
    primaryClass = "hep-th",
    reportNumber = "CERN-TH-2016-050",
    doi = "10.1007/JHEP08(2016)036",
    journal = "JHEP",
    volume = "08",
    pages = "036",
    year = "2016"
}

@article{Li:2017ddj,
    author = "Li, Daliang and Meltzer, David and Stergiou, Andreas",
    title = "{Bootstrapping mixed correlators in 4D $ \mathcal{N} $ = 1 SCFTs}",
    eprint = "1702.00404",
    archivePrefix = "arXiv",
    primaryClass = "hep-th",
    reportNumber = "CERN-TH-2017-024",
    doi = "10.1007/JHEP07(2017)029",
    journal = "JHEP",
    volume = "07",
    pages = "029",
    year = "2017"
}

@article{Behan:2016dtz,
    author = "Behan, Connor",
    title = "{PyCFTBoot: A flexible interface for the conformal bootstrap}",
    eprint = "1602.02810",
    archivePrefix = "arXiv",
    primaryClass = "hep-th",
    doi = "10.4208/cicp.OA-2016-0107",
    journal = "Commun. Comput. Phys.",
    volume = "22",
    number = "1",
    pages = "1--38",
    year = "2017"
}

@article{Fitzpatrick:2011dm,
    author = "Fitzpatrick, A. Liam and Kaplan, Jared",
    title = "{Unitarity and the Holographic S-Matrix}",
    eprint = "1112.4845",
    archivePrefix = "arXiv",
    primaryClass = "hep-th",
    reportNumber = "SLAC-PUB-14979",
    doi = "10.1007/JHEP10(2012)032",
    journal = "JHEP",
    volume = "10",
    pages = "032",
    year = "2012"
}

@article{Mukhametzhanov:2018zja,
    author = "Mukhametzhanov, Baur and Zhiboedov, Alexander",
    title = "{Analytic Euclidean Bootstrap}",
    eprint = "1808.03212",
    archivePrefix = "arXiv",
    primaryClass = "hep-th",
    doi = "10.1007/JHEP10(2019)270",
    journal = "JHEP",
    volume = "10",
    pages = "270",
    year = "2019"
}

@article{Ghosh:2023onl,
    author = "Ghosh, Kausik and Zheng, Zechuan",
    title = "{Numerical conformal bootstrap with analytic functionals and outer approximation}",
    eprint = "2307.11144",
    archivePrefix = "arXiv",
    primaryClass = "hep-th",
    doi = "10.1007/JHEP09(2024)143",
    journal = "JHEP",
    volume = "09",
    pages = "143",
    year = "2024"
}

@article{Paulos:2019gtx,
    author = "Paulos, Miguel F.",
    title = "{Analytic functional bootstrap for CFTs in $d > 1$}",
    eprint = "1910.08563",
    archivePrefix = "arXiv",
    primaryClass = "hep-th",
    doi = "10.1007/JHEP04(2020)093",
    journal = "JHEP",
    volume = "04",
    pages = "093",
    year = "2020"
}

@article{Mazac:2018mdx,
    author = "Mazac, Dalimil and Paulos, Miguel F.",
    title = "{The analytic functional bootstrap. Part I: 1D CFTs and 2D S-matrices}",
    eprint = "1803.10233",
    archivePrefix = "arXiv",
    primaryClass = "hep-th",
    reportNumber = "LPTENS/18/05, LPTENS-18-05",
    doi = "10.1007/JHEP02(2019)162",
    journal = "JHEP",
    volume = "02",
    pages = "162",
    year = "2019"
}

@article{Mazac:2018ycv,
    author = "Mazac, Dalimil and Paulos, Miguel F.",
    title = "{The analytic functional bootstrap. Part II. Natural bases for the crossing equation}",
    eprint = "1811.10646",
    archivePrefix = "arXiv",
    primaryClass = "hep-th",
    doi = "10.1007/JHEP02(2019)163",
    journal = "JHEP",
    volume = "02",
    pages = "163",
    year = "2019"
}

@article{Mazac:2016qev,
    author = "Mazac, Dalimil",
    title = "{Analytic bounds and emergence of AdS$_{2}$ physics from the conformal bootstrap}",
    eprint = "1611.10060",
    archivePrefix = "arXiv",
    primaryClass = "hep-th",
    doi = "10.1007/JHEP04(2017)146",
    journal = "JHEP",
    volume = "04",
    pages = "146",
    year = "2017"
}

@article{Caron-Huot:2022sdy,
    author = "Caron-Huot, Simon and Coronado, Frank and Trinh, Anh-Khoi and Zahraee, Zahra",
    title = "{Bootstrapping $ \mathcal{N} $ = 4 sYM correlators using integrability}",
    eprint = "2207.01615",
    archivePrefix = "arXiv",
    primaryClass = "hep-th",
    doi = "10.1007/JHEP02(2023)083",
    journal = "JHEP",
    volume = "02",
    pages = "083",
    year = "2023"
}

@article{Chang:2025toappear,
    author = "Chang, Cyuan-Han and Dommes, Vasiliy and Erramilli, Rajeev S. and Homrich, Alexandre and Kravchuk, Petr and Mitchell, Matthew S. and Poland, David and Simmons-Duffin, David",
    title = "{to appear}",
}

@article{Caron-Huot:2020adz,
    author = "Caron-Huot, Simon and Mazac, Dalimil and Rastelli, Leonardo and Simmons-Duffin, David",
    title = "{Dispersive CFT Sum Rules}",
    eprint = "2008.04931",
    archivePrefix = "arXiv",
    primaryClass = "hep-th",
    doi = "10.1007/JHEP05(2021)243",
    journal = "JHEP",
    volume = "05",
    pages = "243",
    year = "2021"
}

@article{Penedones:2019tng,
    author = "Penedones, Joao and Silva, Joao A. and Zhiboedov, Alexander",
    title = "{Nonperturbative Mellin Amplitudes: Existence, Properties, Applications}",
    eprint = "1912.11100",
    archivePrefix = "arXiv",
    primaryClass = "hep-th",
    reportNumber = "CERN-TH-2019-230",
    doi = "10.1007/JHEP08(2020)031",
    journal = "JHEP",
    volume = "08",
    pages = "031",
    year = "2020"
}

@article{Carmi:2020ekr,
    author = "Carmi, Dean and Penedones, Joao and Silva, Joao A. and Zhiboedov, Alexander",
    title = "{Applications of dispersive sum rules: $\varepsilon$-expansion and holography}",
    eprint = "2009.13506",
    archivePrefix = "arXiv",
    primaryClass = "hep-th",
    reportNumber = "CERN-TH-2020-162",
    doi = "10.21468/SciPostPhys.10.6.145",
    journal = "SciPost Phys.",
    volume = "10",
    number = "6",
    pages = "145",
    year = "2021"
}

@article{Mazac:2019shk,
    author = "Maz{\'a}{\v{c}}, Dalimil and Rastelli, Leonardo and Zhou, Xinan",
    title = "{A basis of analytic functionals for CFTs in general dimension}",
    eprint = "1910.12855",
    archivePrefix = "arXiv",
    primaryClass = "hep-th",
    doi = "10.1007/JHEP08(2021)140",
    journal = "JHEP",
    volume = "08",
    pages = "140",
    year = "2021"
}

@article{Poland:2011ey,
    author = "Poland, David and Simmons-Duffin, David and Vichi, Alessandro",
    title = "{Carving Out the Space of 4D CFTs}",
    eprint = "1109.5176",
    archivePrefix = "arXiv",
    primaryClass = "hep-th",
    doi = "10.1007/JHEP05(2012)110",
    journal = "JHEP",
    volume = "05",
    pages = "110",
    year = "2012"
}

@article{Kos:2014bka,
    author = "Kos, Filip and Poland, David and Simmons-Duffin, David",
    title = "{Bootstrapping Mixed Correlators in the 3D Ising Model}",
    eprint = "1406.4858",
    archivePrefix = "arXiv",
    primaryClass = "hep-th",
    doi = "10.1007/JHEP11(2014)109",
    journal = "JHEP",
    volume = "11",
    pages = "109",
    year = "2014"
}

@article{Simmons-Duffin:2015qma,
    author = "Simmons-Duffin, David",
    title = "{A Semidefinite Program Solver for the Conformal Bootstrap}",
    eprint = "1502.02033",
    archivePrefix = "arXiv",
    primaryClass = "hep-th",
    doi = "10.1007/JHEP06(2015)174",
    journal = "JHEP",
    volume = "06",
    pages = "174",
    year = "2015"
}

@article{Landry:2019qug,
    author = "Landry, Walter and Simmons-Duffin, David",
    title = "{Scaling the semidefinite program solver SDPB}",
    eprint = "1909.09745",
    archivePrefix = "arXiv",
    primaryClass = "hep-th",
    reportNumber = "CALT-TH 2019-038",
    month = "9",
    year = "2019"
}

@article{Rattazzi:2008pe,
    author = "Rattazzi, Riccardo and Rychkov, Vyacheslav S. and Tonni, Erik and Vichi, Alessandro",
    title = "{Bounding scalar operator dimensions in 4D CFT}",
    eprint = "0807.0004",
    archivePrefix = "arXiv",
    primaryClass = "hep-th",
    doi = "10.1088/1126-6708/2008/12/031",
    journal = "JHEP",
    volume = "12",
    pages = "031",
    year = "2008"
}

@article{Rattazzi:2010gj,
    author = "Rattazzi, Riccardo and Rychkov, Slava and Vichi, Alessandro",
    title = "{Central Charge Bounds in 4D Conformal Field Theory}",
    eprint = "1009.2725",
    archivePrefix = "arXiv",
    primaryClass = "hep-th",
    reportNumber = "LPTENS-10-35",
    doi = "10.1103/PhysRevD.83.046011",
    journal = "Phys. Rev. D",
    volume = "83",
    pages = "046011",
    year = "2011"
}

@article{Poland:2010wg,
    author = "Poland, David and Simmons-Duffin, David",
    title = "{Bounds on 4D Conformal and Superconformal Field Theories}",
    eprint = "1009.2087",
    archivePrefix = "arXiv",
    primaryClass = "hep-th",
    doi = "10.1007/JHEP05(2011)017",
    journal = "JHEP",
    volume = "05",
    pages = "017",
    year = "2011"
}

@article{Ber31,
  author  = {Bernstein, Serge},
  title   = {{Sur la limitation des valeurs d'un polyn\^ome $P_n(x)$ de degr\'e $n$ sur tout un segment par ses valeurs en $(n+1)$ points du segment}},
  journal = {Izv. Akad. Nauk SSSR},
  year    = {1931},
  volume  = {8},
  pages   = {1025--1050}
}

@book{Walsh,
  author    = {Walsh, J. L.},
  title     = {Approximation by Polynomials in the Complex Domain},
  series    = {Mémorial des sciences mathématiques},
  number    = {73},
  year      = {1935},
  publisher = {Gauthier-Villars, Éditeur, Paris},
  address   = {Paris},
  pages     = {72},
  note      = {Published under the patronage of the Académie des sciences de Paris},
}
\end{document}